\documentclass[12pt]{article}

\usepackage{graphicx}
\usepackage{epstopdf}
\usepackage[body={17.5cm, 22cm},right=2.2cm]{geometry}
\usepackage{amssymb}
\usepackage{amsmath}

\newcommand\ee{\end{equation}}
\newcommand\be{\begin{equation}}
\newcommand\eea{\end{eqnarray}}
\newcommand\bea{\begin{eqnarray}}
\newcommand{\sfrac}[2]{{\textstyle\frac{#1}{#2}}}
\newcommand\Lc{\Lambda^3}
\newcommand\di{\partial}
\newcommand\de{\partial}

\begin{document}

\setcounter{page}{0}
\thispagestyle{empty}
\begin{flushright}
CERN-PH-TH/2006-033\\
HUTP-06/A0005
\end{flushright}
\begin{center}

~

\vspace{1.cm}

{\Large
\bf{
Causality, Analyticity and an\\[0.4cm]  IR Obstruction to UV Completion
}}\\[1cm]
%
%
%
{\large Allan Adams$^{\rm a}$, Nima Arkani-Hamed$^{\rm a}$, Sergei Dubovsky$^{\rm a,b, c}$,\\ [0.2cm]Alberto Nicolis$^{\rm a}$, Riccardo Rattazzi$^{\rm b}$\footnote{On leave from INFN, Pisa, Italy.}}
\\[0.5cm]

{\small \textit{$^{\rm a}$ Jefferson Physical Laboratory, \\ Harvard University, Cambridge, MA 02138, USA}}

\vspace{.2cm}
{\small \textit{$^{\rm b}$ CERN Theory Division, CH-1211 Geneva 23, Switzerland}}

\vspace{.2cm}
{\small \textit{$^{\rm c}$ Institute for Nuclear Research of the Russian Academy of Sciences, \\
        60th October Anniversary Prospect, 7a, 117312 Moscow, Russia}}

\end{center}

\vspace{.8cm}
\vspace{0.3cm} {\small \noindent \textbf{Abstract} \\[0.3cm]
\noindent
We argue that certain apparently consistent low-energy effective
field theories described by local, Lorentz-invariant Lagrangians,
secretly exhibit macroscopic non-locality and cannot be embedded
in any UV theory whose $S$-matrix satisfies canonical analyticity constraints.
The obstruction involves the signs of a set of leading irrelevant
operators, which must be strictly positive to ensure UV analyticity.
An IR manifestation of this restriction is that the ``wrong'' signs
lead to superluminal fluctuations around non-trivial backgrounds, making
it impossible to define local, causal evolution, and implying a surprising IR breakdown of the effective theory.
Such effective theories can not arise in quantum field theories or weakly
coupled string theories, whose $S$-matrices satisfy the usual analyticity
properties.  This conclusion applies to the DGP
brane-world model modifying gravity in the IR, giving a simple
explanation for the difficulty of embedding this model into
controlled stringy backgrounds, and to models of
electroweak symmetry breaking that predict negative anomalous
quartic couplings for the $W$ and $Z$.  Conversely, any experimental
support for the DGP model, or measured negative signs for anomalous
quartic gauge boson couplings at future accelerators, would
constitute direct evidence for the existence of superluminality
and macroscopic non-locality unlike anything previously seen in
physics, and almost incidentally falsify both local quantum field theory
and perturbative string theory.}

\vfill

\setcounter{footnote}{0}
\section{Introduction}

Can every low-energy effective theory be UV completed into a full theory? To a string theorist in 1985, the answer to this question would have been a resounding ``no."  The hope was that the consistency conditions on a full theory of quantum gravity would be so strong as to more or less uniquely single out the standard model coupled to GR as the unique low-energy effective theory, and that the infinite number of other possible effective theories simply couldn't be extended to a
full theory. In support of this view, the early study of perturbative heterotic strings
yielded many constraints on the properties of the low-energy theory invisible
to the effective field theorist. For instance, the rank of the gauge group was
restricted to be smaller than 22.

With the discovery of D-branes and the duality revolution, these
constraints appear to have evaporated, leaving us with a continuous
infinity of consistent supersymmetric theories coupled to gravity
and very likely a huge discretum of non-supersymmetric vacua
\cite{Shamit}. If the low-energy theory describing our universe is
not unique but merely one point in a vast landscape of vacua of
the underlying theory, then the properties of our
vacuum---such as the values of the dimensionless couplings of
the standard model---are unlikely to be tied to the structure of
the fundamental theory in any direct way, reducing the detailed
study of its particle-physical properties to a problem of only
parochial interest. This situation is not without its consolations.
With a vast landscape of vacua,  seemingly intractable fine-tuning
puzzles such as the cosmological constant problem \cite{weinberg},
and perhaps even the hierarchy problem \cite{don}, can be solved by
being demoted from fundamental questions to environmental ones,
suggesting new models for particle physics \cite{split}.

Given these developments, it is worth asking again: can every effective field theory be UV completed?
The evidence for an enormous landscape of vacua in string theory
certainly encourages this point of view---if even the consistency
conditions on quantum gravity leave room for huge numbers of
consistent theories, surely any consistent model can be embedded
somewhere in the landscape. Much of the activity in model-building
in the last five years has implicitly taken this point of view,
constructing interesting theories purely from the bottom-up with no
obvious embedding into any microscopic theory. This has been
particularly true in the context of attempts to modify gravity in
the infrared, including most notably the Dvali-Gabadadze-Porrati
model \cite{DGP} and more recent ideas on Higgs phases of gravity
\cite{ghost,Rub,Dub}.

In this note, we wish to argue that the pendulum has swung too
far in the ``anything goes" direction.  Using simple and familiar
arguments, we will show that some apparently perfectly sensible
low-enegy effective field theories governed by local,
Lorentz-invariant Lagrangians, are secretly non-local, do not admit
any Lorentz-invariant notion of causality, and are incompatible with a
microscopic $S$-matrix satisfying the usual analyticity conditions.
The consistency condition we identify is that the signs of certain
higher-dimensional operators in any non-trivial effective theory
must all be strictly positive. The inconsistency of theories which
violate this positivity condition has both UV and IR avatars.

The IR face of the problem is that, for the ``wrong" sign of these
operators, small fluctuations around translationally invariant
backgrounds propagate superluminally, making it
impossible to define a Lorentz-invariant time-ordering of events.
Moreover, in general backgrounds, the equation of motion can
degenerate on macroscopic scales to a non-local constraint
equation whose solutions are UV-dominated.
Thus, while these theories are local in the sense that the field
equations derive from a strictly local Lagrangian, and Lorentz-invariant
in the sense that Lorentz  transforms of solutions to the field equations
are again solutions, the macroscopic IR physics of this theory is
neither Lorentz-invariant nor local.

The UV face of the problem is also easy to discern: assuming that UV
scattering amplitudes satisfy the usual analyticity conditions,
dispersion relations and unitarity immediately imply a host of
constraints on {\it low energy} amplitudes. One particular such
constraint is that that the leading low energy forward scattering
amplitude must be non-negative, yielding the same positivity
condition on the higher-derivative interactions as the
superluminality constraint. Of course the fact that analyticity and
unitarity imply positivity constraints is very well known, and the
connection of analyticity to causality is an ancient one.

We will focus on models in which the UV cutoff is far beneath the
(four-dimensional) Planck scale, so gravity in unimportant, though
we will also make some comments about gravitational theories. Our
work thus complements the intrinsically gravitational limitations on
effective field theories recently discussed in \cite{weakg,swamp}.

Of course, local quantum field theories have a Lorentz-invariant
notion of causality and satisfy the usual $S$-matrix axioms,
so any effective field theory which violates our positivity conditions
cannot be UV completed into a local QFT.  Significantly,
since weakly coupled string amplitudes satisfy the same
analyticity properties as amplitudes in local quantum field theories---
indeed, the Veneziano amplitude arose from $S$-matrix theory---
the same argument applies to weakly coupled strings.
Thus, while string theory is certainly non-local in many crucial ways,
the effective field theories arising from string theory are in this
precise sense just as local as those deriving from local quantum
field theory, and satisfy the same positivity constraints.

Positivity thus provides a tool for identifying what physics can and
cannot arise in the landscape. Perhaps surprisingly, the tool is a
powerful one.  For example, it is easy to check that the DGP model
violates positivity, providing a simple explanation for why this
model has so far resisted an embedding in controlled weakly coupled
string backgrounds. Similarly, certain 4-derivative terms in the
chiral Lagrangian are constrained to be positive, implying for
example that the electroweak chiral Lagrangian cannot be UV
completed unless the anomalous quartic gauge boson couplings are
positive.

The flipside of this argument is that any experimental evidence of a
violation of these positivity constraints would signal a crisis for
the usual rules of macroscopic locality, causality and analyticity,
and, almost incidentally, falsify perturbative string theory. For
example, the DGP model makes precise predictions for deviations in
the moon's orbit that will be checked by laser lunar ranging
experiments \cite{moon}. If these deviations are seen and other
pieces of experimental evidence supporting the DGP effective theory
are gathered, we would also have evidence for parametrically fast
superluminal signal propagation and macroscopic violation of
locality, as well as a non-analytic $S$-matrix, unlike anything
previously seen in physics. The same conclusion holds if future
colliders indicate evidence for {\it negative} anomalous quartic
gauge boson couplings. Experimental evidence for either of these
theories would therefore clearly disprove some of our fundamental
assumptions about physics.


\section{Examples}

Let's begin with some examples of the apparently consistent
low-energy effective theories we will constrain. Of course we should
be precise about what we mean by a consistent effective
theory---loosely it should have stable vacuum, no anomalies and so
on, but most precisely, a consistent effective field theory is just
one that produces an exactly unitary $S$-matrix for particle
scattering at energies beneath some scale $\Lambda$.

Consider the theory of a
single $U(1)$ gauge field. The leading interactions in this theory
are irrelevant operators,
\begin{equation} \label{EH}
{\cal L} = -\frac{1}{4} F_{\mu \nu} F^{\mu \nu} + \frac{c_1}{\Lambda^4}
(F_{\mu \nu} F^{\mu \nu})^2 + \frac{c_2}{\Lambda^4} (F_{\mu \nu}
\tilde{F}^{\mu \nu})^2 + \dots \: ,
\end{equation}
with $\Lambda$ some mass scale and $c_{1,2}$ dimensionless coefficients.
As another example, consider a massless scalar field $\pi$ with a
shift symmetry $\pi \to \pi + {\rm const}$. Again the leading
interactions are irrelevant,
\begin{equation} \label{goldstone}
{\cal L} = \partial^\mu \pi \partial_\mu \pi   + \frac{c_3}{\Lambda^4}
(\partial_\mu \pi \partial^\mu \pi)^2 + \dots
\end{equation}

As far as an effective field theorist is concerned, the coefficients
$c_{1,2,3}$ are completely arbitrary numbers. Whatever the $c_i$
are, they can give the leading amplitudes in an exactly unitary
$S$-matrix at energies far beneath $\Lambda$. Of course the theories
are non-renormalizable so an infinite tower of higher operators must
be included, nonetheless there is a systematic expansion for the
scattering amplitudes in powers of $(E/\Lambda)$ which is unitary to
all orders in this ratio. However, we claim that in any UV
completion which respects the usual axioms of $S$-matrix theory, the
$c_i$ are forced to be {\it positive}

\begin{equation}
c_{i} > 0 \; .
\end{equation}

It is easy to check that indeed these coefficients are positive in
all familiar UV completions of these models. For instance, the
Euler-Heisenberg Lagrangian for QED, arising from integrating out
electrons at 1-loop, indeed generates $c_{1,2} > 0$. Analogously, we can identify $\pi$ as a Goldstone boson
in a linear sigma model, where $\pi$ and a Higgs
field $h$ are united into a complex scalar field $\Phi$,
\begin{equation}
\Phi = (v + h) e^{i \pi/v} \; ,
\end{equation}
with a potential $V(|\Phi|) = \lambda(|\Phi|^2 - v^2)^2$. The action for $\pi, h$ at
tree-level is
\begin{equation}
{\cal L} = \Big(1 + \frac{h}{v}\Big)^2 (\partial \pi)^2 + (\partial h)^2 - M_h^2
h^2 - \dots
\end{equation}
Integrating out $h$ at tree-level yields the quartic term
\begin{equation}
{\cal L}_{\rm eff} =\frac{\lambda}{M_h^4}
(\partial \pi)^4 + \dots
\end{equation}
which has the claimed positive sign.

Another example involves the fluctuations of a brane in an
extra dimension, given by a field $y(x)$ with the effective
lagrangian
\begin{equation}
{\cal L} = - f^4 \sqrt{1 - (\partial y)^2} = f^4 \Big[- 1 +
\frac{(\partial y)^2}{2} + \frac{(\partial y)^4}{8} + \dots
\Big] \; .
\end{equation}
Again we find the correct sign. Related to this, the Born-Infeld
action for a $U(1)$ gauge field localized to a D-brane also
gives the correct sign for all $F^4$ terms.

There are also other simple 1-loop checks.
For example, imagine coupling $N$ fermions to $\Phi$ in our UV linear sigma model; for sufficietly large $N$, 1-loop
effects can dominate over the tree terms coming from integrating out the Higgs. For instance, consider integrating
out a higgsed fermion. Grouping two
Weyl fermions $\psi,\psi^c$ with charges $\pm 1$
into a Dirac spinor $\Psi$, the effective Lagrangian is
\begin{equation}
\bar{\Psi} \Big[i \gamma^\mu \big(\partial_\mu + i \frac{\partial_\mu
\pi}{v} \, \gamma^5 \big) - M_\Psi \Big]\Psi \; .
\end{equation}
At 1-loop, we generate an effective quartic interaction
\begin{equation}
{\cal L}_{\rm eff} = \frac{1}{48 \pi^2 v^4} (\partial \pi)^4 +
\dots \; ,
\end{equation}
resulting again in a positive leading irrelevant operator.

Note that the positivity constraints we are talking about are not
directly related to other familiar positivity constraints that
follow from vacuum stability. We know for instance that kinetic
terms are forced to be positive, and that $m^2 \phi^2$ and $\lambda
\phi^4$ couplings must also be positive. In all these cases, the
``wrong" signs are associated with a  clear instability already
visible in the low-energy theory. Related to this, the euclidean
path integrals for such theories are not well-defined, having
non-positive-definite euclidean actions.

By contrast, the ``wrong" sign for the leading derivative interactions (such as the  $(\partial \pi)^4$ terms above) are not associated with any energetic instabilities in the low-energy vacuum: the correct sign of the
kinetic terms guarantee that all gradient energies are positive,
with the terms proportional to the $c_i$ giving only small corrections
within the effective theory.  Indeed, even if the leading irrelevant operators---the only ones to which our constraints apply---have the ``wrong'' sign, higher order terms can
ensure the positivity of energy (at least classically), e.g.~higher powers of $(\partial \pi)^2$.
Related to this, the euclidean path integrals in theories with ``wrong'' signs do not exhibit any obvious pathologies. Of course this non-renormalizable theory must be treated using the standard ideas of effective field theory, but the healthy euclidean formulation at least perturbatively guarantees a unitary low-energy $S$-matrix when we continue back to Minkowski space.


\section{Signs and Superluminality}

If models with the ``wrong''
signs have stable, Lorentz-invariant vacuua with perfectly sensible and unitary
perturbative $S$-matrices, why don't they arise as the low-energy limit of any familiar UV-complete theories? As we will see, while the trivial vacua of such theories are well-behaved, the speed
of fluctuations around {\it non}-trivial
backgrounds depend critically on these signs, with the
``wrong'' signs leading to superluminal propagation in generic backgrounds.  This in turn leads to familiar conflicts with causality and locality which are not present in any microscopically local quantum field or perturbative string theory.  Exactly how this conflict arises turns out to be an illuminating question.

Let's begin by establishing the connection between positivity-violating irrelevant leading interactions and superluminality in non-trivial backgrounds.  Suppose we expand the effective theory around some non-trivial translationally invariant solution of the field equations.
As long as the background field is sufficiently small, the effective field theory remains valid.
Translational invariance ensures that small fluctuations satisfy a simple dispersion relation, $\omega^2 = v^2 (\hat k) \, |\vec{k}|^2$,  with the velocity $v(\hat k)$ determined by the higher-dimension operators in the lagrangian.  The crucial insight is that whether fluctuations travel slower or faster than light depends entirely on the {\it signs} of the leading irrelevant interactions.

Let's see how this works in an explicit example.
Consider our Goldstone model expanded around the solution $\di_\mu \pi_{0} = C_\mu$, where $C_\mu$ is a constant vector.
The linearized equation of motion for fluctuations $\varphi \equiv \pi -\pi_0$ around this background is
\be \label{fluct_goldstone}
\left[ \eta^{\mu\nu} +4 \frac{c_3}{\Lambda^4} \, C^\mu C^\nu + \dots  \right]
\di_\mu\di_\nu \varphi = 0 \; .
\ee
Within the regime of validity of the effective theory, $C_\mu C^\mu \ll \Lambda^4$, all higher dimension interactions are negligible - all that matters is the {\it leading} interaction, $c_{3}$.  Expanding in plane waves, this reads
\be
k^\mu k_\mu + 4 \frac{c_3}{\Lambda^4} (C\cdot k)^{2}=0 \; .
\ee
Since $(C\cdot k)^{2} \ge 0$, {\it the absence of superluminal excitations requires that the coefficient $c_{3}$ is positive.}

The case of the electromagnetic field is slightly more involved---the speed of fluctuations around non-trivial backgrounds now depends on both momentum {\it and} polarization $\epsilon_{\mu}$, and thus on {\it both} of the leading interactions in the Lagrangian,
\be
k^\mu k_\mu ~+ 32 \frac{c_1}{\Lambda^4} (F^{\mu\nu}k_{\mu}\epsilon_{\nu})^{2} ~+~  32 \frac{c_2}{\Lambda^4} (\tilde{F}^{\mu\nu}k_{\mu}\epsilon_{\nu})^{2}=0 \, ,
\ee
but the conclusion is completely analogous: {\em there exist no superluminal excitations iff the coefficients $c_{1,2}$ are both positive}.  Note that these conclusions hold independently of the particular background field one turns on.
Note too that even when the shift in the speed of propagation is very small, $v-1 \sim \frac{C^{2}}{\Lambda^4}\ll1$, it can easily be measured in the low-energy effective theory by allowing signals to propagate over large distances.   It is interesting to note that in the case of open strings on D-branes, which are governed by a BI Lagrangian of the form (1), the speed of propagation in the presence of a background fieldstrength $(F+B)\neq 0$ can be computed exactly in terms of the so-called "open string metric" and is always slower than the speed of light -- which is to say, this appearance of the BI Lagrangian in string theory satisfies positivity, with $c_{1,2}>0$.

%
%

At this point all the problems usually associated with superluminality---the ability to send signals back in time, closed timelike curves, etc.---rear their heads.  On the other hand, such effects are appearing within a theory governed by a local Lorentz-invariant lagrangian, a hyperbolic equation of motion and a perfectly stable vacuum.  It is thus instructive to work through the physical consequences of this kind of superluminality and understand exactly when and why these theories run into trouble.

\subsection{The Trouble with Lorentz Invariance}

That the effective Lagrangian is Lorentz-invariance ensures that Lorentz transforms of solutions to the field equations are again solutions to the field equations.   It does not, however, ensure that all inertial frames are on an even footing.
Consider for example the equation of motion for fluctuations $\varphi$ around translationally-invariant backgrounds of our Goldstone model,
$$
\partial_{t}^2 \varphi - v^2 \partial_{i}^2 \varphi = 0 \, ,
$$
where $v^2  \simeq 1  -\frac{4c_3}{\Lambda^4}C^{2} $ is the velocity of propagation.  This has oscillatory solutions propagating in all directions, e.g.~$\varphi=f(x\pm vt)$. Upon boosting in, say, the $\hat{x}$ direction, the equation of motion becomes
$$
(1-v^2\beta^2) ~ \partial_t^2 \varphi  ~ +2\beta(1-v^2)~\partial_t\partial_x\varphi  ~- (v^2-\beta^2)~\partial_x^2 \varphi
~- v^2~\partial_{\perp}^2 \varphi = 0 \, .
$$
whose solutions, e.g.~$\varphi=f(x\pm\frac{v-\beta}{1\mp v\beta}t)$, are the Lorentz boost of the original solutions.  So far so good.  However, if $v^2>1$, there exists a frame ($\beta=1/v$) in which the coefficient of $\partial_t^2 \varphi$ vanishes, $\varphi$ propagates instantaneously and the equation of motion becomes a non-dynamical constraint.  In this frame it is simply impossible to set up an initial value problem to evolve the field from Cauchy slice to Cauchy slice\footnote{Notice that we are
dealing with tiny superluminal shifts in the dispersion relation, so
we need huge boost velocities to observe these effects,
requiring both $\dot \pi$ and $\nabla \pi$ to be of order
$\Lambda$; however, since the Lorentz invariant combination $\di_\mu
\pi \di^\mu \pi$ remains tiny, the description of the system in
terms of the effective theory remains valid for all observers,
ensuring that these effects obtain well within the domain
of validity of effective field theory.}.
When $\beta > 1/v$, the equation of motion is again perfectly dynamical and can certainly be integrated---however, oscillatory solutions to these equations move only in the {\it positive} $x$ direction, while modes in other directions
may be exponentially growing or decaying.  What's going on?  How is it possible that what looks like a stable system in one frame looks horribly unstable in another?

The point is that what look like perfectly natural initial conditions for a superluminal mode in one frame look like horribly fine-tuned conditions in another.  Indeed, the time-ordering of events connected by propagating fluctuations is not Lorentz-invariant.
Observers in relative motion will thus disagree rather dramatically about what constitutes a sensible set of initial conditions to propagate with their equations of motion---initial conditions that to one observer look like turning on a localized source at some unremarkable point in spacetime will appear to the other as a bewildering array of fluctuations incident from past infinity which conspire miraculously to annihilate what the original observer wanted to call the localized source.  Said differently, the retarded Green function in one frame is a mixture of advanced and retarded Green functions in another frame.    {\it Fixing initial conditions on past infinity thus explicitly breaks Lorentz-invariance.}  In order for the theory to be predictive, we must choose a frame in which to define retarded Green functions.  In sufficiently well-behaved backgrouds, there is a particularly natural choice of frame, that in which such conspiracies do {\it not} appear.

Returning to our question of stability vs instability, consider a solution in the highly boosted frame in which we turn on a localized source for one of the unstable excitations.  A Lorentz boost unambiguously maps this to a solution in the stable unboosted frame.  The crucial point is that the resulting configuration does {\it not} look like a small fluctuation sourced by a local source---indeed, these are explicitly stable according to the equation of motion---but rather involves turning on initial conditions at a fixed time which vary exponentially in {\it space}, {\it along} the slice.  These do not represent instabilities in any usual sense; they simply represent initial conditions which we would normally rule out as unphysical.
By the same token, a localized fluctuation which remains everywhere bounded and oscillatory in the original frame transforms into a miraculous conspiracy in the initial conditions that prevents the apparently unstable mode from turning on and growing.  Crucially, this never happens in theories with null or timelike propagation, in which Lorentz transformations carry sensible initial conditions to sensible initial conditions.

\begin{figure}[b!]
\begin{center}
\includegraphics[width=13cm]{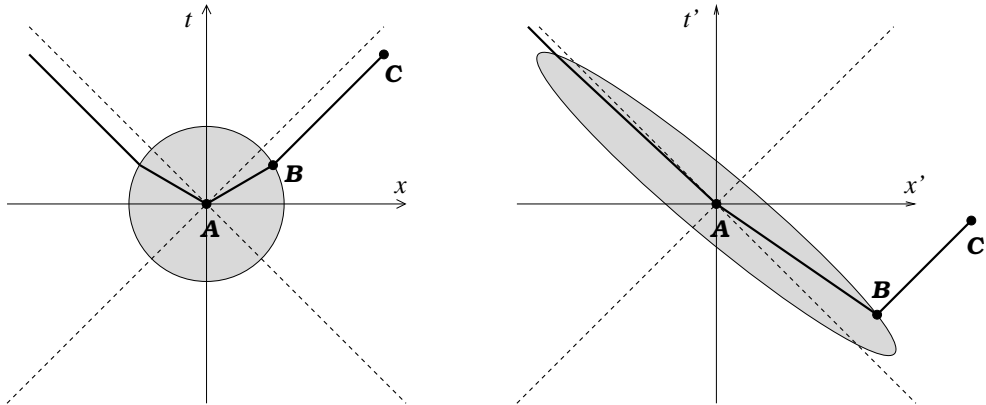}
\end{center}
\caption{\label{blob}
\footnotesize Bubbles of non-trivial vacua, $\pi = C_\mu x^\mu$, in our Goldstone model with $c_{3}<0$.  (a) In the rest frame of the bubble, $C_{\mu}=(C,0,0,0)$.  The solid lines denote the causal cone inside of which small fluctuations are constrained to propagate.  (b) The same system in a boosted frame in which the bubble moves with a large velocity in the positive $x'$ direction.  For sufficiently large boosts, the causal cone dips below horizontal, and small fluctuations are only seen to propagate to the left with a different temporal ordering than in the unboosted frame.}
\end{figure}

It is enlightening to run through the above logic in translationally non-invariant backgrounds.
Consider again the Goldstone model with ``wrong'' sign, $c_{3}<0$, and imagine building, by suitable arrangement of sources, a finite-sized bubble of $\di_\mu \pi = C_\mu$ condensate localized in space and time.  Let's begin in the rest frame of the condensate, in which $C_{\mu}=(C,0,0,0)$.  Outside the bubble, in the trivial $\pi=0$ vacuum, fluctuations of $\pi$ satisfy the massless wave equation and propagate along null rays.  Inside, however, fluctuations move with velocity
$v^2  \simeq 1  -\frac{4c_3}{\Lambda^4}C^{2}$
and thus propagate not along the light cone but along a ``causal'' cone defined by the effective metric $G_{\mu\nu}=\eta_{\mu \nu} + 4\frac{c_3}{\Lambda^4} \partial_\mu \pi \partial_\nu \pi$. When $c_{3}<0$, this cone is broader than the light cone and fluctuations propagate ever so slightly superluminally (see fig.~\ref{blob}a).  However, since fluctuations always propagate forward in time, setting up and solving the Cauchy problem in this background is still no problem.

As above, when $c_{3}<0$ it is possible for the coefficient of the $\partial_t^2 \varphi $ term in the equation of motion of a rapidly moving observer to vanish (see fig.~\ref{blob}b).
Inside the bubble the coefficient of $\partial_t^2 \varphi$ in the
equation of motion is negative, while outside it is positive---somewhere along the boundary of the bubble, then, the coefficient must pass through zero, at which point the equation of motion becomes again a constraint.   Thus, in any frame in which the causal cone deep inside the bubble dips below the horizontal, the bubble has a closed shell on which evolution from timeslice to timeslice cannot be prescribed by local hamiltonian flow.   This in fact helps explain the peculiar phenomena seen by this boosted observer.
Consider the sequence of events depicted in fig.~\ref{blob}. An observer in the rest
frame of the bubble sends a superluminal fluctuation from a point, $A$, deep inside the bubble to a point, $B$, on the boundary at which the wave exits the bubble, proceeding at the speed of light to a distant point, $C$. In a highly boosted frame,
the sequence of events will have $B$ happening before $A$ or $C$. How is
this possible? The resolution is that the coefficient of
$\partial_t^2 \varphi$ vanishes at $B$, so the evolution of $\varphi$ at $B$
can't be predicted from local measurements; instead, a constraint requires the spontaneous
appearance of two $\varphi$ excitation just inside and outside the
bubble, which then continue forwards in time to $A$ and $C$.

This is not something with which we are familiar, and makes it seem unlikely any  Lorentz invariant $S$-matrix exists within such theories.  Indeed, the existence of a prefered class of frames---those in which the field equations do not degenerate to constraint equations---suggests that the Lorentz invariance of the classical Lagrangian is physically irrelevant, and raises doubts about the possibility of embedding such effective theories in UV-complete theories which respect microscopic Lorentz invariance and locality.  Notice that
systems with superluminal propagation are in this sense somewhat analogous to Lorentz invariant field theories
with ghosts, of which no sense can be made unless Lorentz-invariance is explicitly broken.
This is because boost-invariance makes the rate of decay of the vacuum by ghost emission formally infinite---only if
Lorentz-invariance is not a symmetry of the theory can the decay rate be made finite.
In such systems, however, Lorentz-invariance can only arise as an accidental symmetry.

\subsection{Global Problems with Causality}

\begin{figure}[t!]
\begin{center}
\includegraphics[width=10cm]{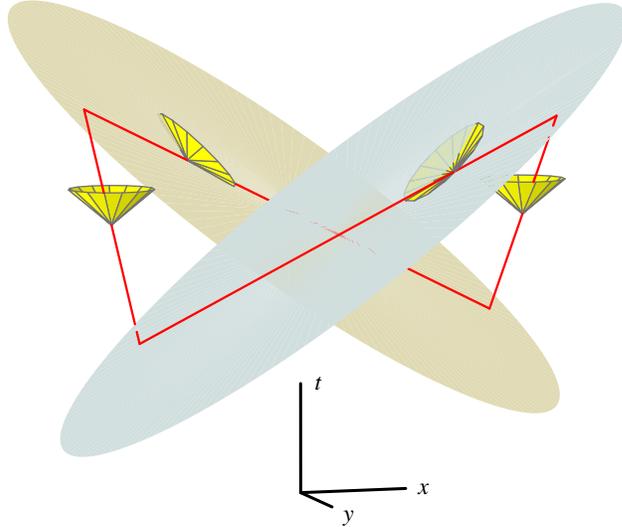}
\end{center}
\caption{\label{ctc}
{\footnotesize  Two finite bubbles moving with large opposite velocities in the $x$ direction and separated by a finite distance in the $y$ direction.  The open cones indicate the local causal cones of $\pi$-fluctuations, and the red line the closed trajectory of a series of small fluctuations along these cones.  Such closed time-like trajectories make it clear that no notion of causality or locality survives in a theory which violates positivity.}}
\end{figure}

In the simple system of a single bubble in otherwise empty space, there always exists families of inertial frames in which causality is meaningfully defined. In particular, the co-moving rest frame of the bubble defines a time slicing in this `good' class, so we can simply declare that evolution is to be prescribed in the rest frame of the bubble and translated into other frames by boosting with the spontaneously broken Lorentz generators.  Forward evolution in time in highly boosted frames may look bizarre to a boosted inertial observer, but it is unambiguous.
However, there are always backgrounds in which no global rest frame exists---for example, two bubbles of $\pi$ condensate flying past each other at high velocity and finite impact parameter, as in fig.~\ref{ctc}---so it is far from obvious whether there is {\it any} good notion of causal ordering in these theories.

It is useful to treat this problem with the aid of some formalism. Consider again the wave equation for small fluctuations around a non-trivial background in the Goldstone system,
\begin{equation}
G^{\mu \nu} \partial_\mu \partial_\nu \varphi = 0 \, ,  \qquad G^{\mu \nu} =
\eta^{\mu \nu} + \frac{4 c_3}{\Lambda^4} \partial^\mu \pi
\partial^\nu \pi\, .
\end{equation}
This equation suggests a natural inverse-metric $G^{\mu\nu}$ with which to define $\varphi$ ``lightcones" and time-evolution \footnote{Strictly speaking the interpretation of $G_{\mu\nu}$ as
an effective metric holds only in the geometric optics limit in which
 the  wavelengths are short enough with respect to the distance over which
$G_{\mu\nu}$ itself varies. Anyway, if a pathology arises already in this limit, and we shall see that it does,
we do not need to worry about the case of long wavelengths.}. The  metric $G_{\mu\nu}$ is indeed
what determines the light cone structure within the blobs in Fig.~\ref{blob}.
Now that we have a metric, we can apply the methodology of General Relativity to determine whether
causality is meaningfully defined over our spacetime~\cite{wald}. A first requirement is that the spacetime be
time orientable, meaning that  there should exist a globally defined and non-degenerate timelike vector, $ t^\mu$.
To see that this is the case, note that $G_{\mu\nu}$ is
\be
G_{\mu\nu}=\eta_{\mu\nu}-\frac{4c_3}{\Lambda^4}\partial_\mu\pi\partial_\nu\pi+\dots
\ee
where the dots stand for terms that can be neglected when $\partial_\mu \pi\partial^\mu\pi/\Lambda^4\ll 1$
and the effective field theory surely makes sense. Then, for $c_3<0$, we have $G_{00}>1$ and therefore the
vector $t^\mu=(1,0,0,0)$ is globally defined, non-degenrate and time-like.
The vector $t^\mu$ defines at each
space-time point the direction of time flow. Future directed timelike curves  $x^\mu(\sigma)$ are those defined
by
\be
\dot x^\mu t^\nu G_{\mu\nu} >0\qquad\qquad \dot x^\mu\dot x^\nu G_{\mu\nu}>0\,  .
\ee
The second  condition for  causality  to hold is that there be no closed (future directed) timelike curves
(CTCs). In the presence of CTCs, the $t$ coordinate is not globally defined---it is multiply valued---and
time evolution again becomes a constrained, non-local problem, and causality is lost.

The Goldstone and the Euler-Heisemberg  systems are both time orientable, at least
for backgrounds within the domain of validity of the effective field theory description.
Moreover, for simple backgrounds like the single bubble of Fig.~\ref{blob} it  is also evident that there are
no CTCs, so that a sensible, although not Lorentz invariant, notion of causality exists.
However, in both systems, there exist other backgrounds in which the effective metric $G_{\mu\nu}$ {\it does} admit CTCs, and time evolution can {\it not} be locally defined but must satisfy {\it globally} constraints.

In our Goldstone system, a simple such offending background is given by
two superluminal bubbles flying rapidly past each other, as shown in Fig.~\ref{blob}.  Note that a head on collision
between the two bubbles in the same plane would certainly take us out
of the regime of validity of the effective theory, with $(\partial
\pi)^2$ becoming large in the overlap region. But it is easy to check
that a small separation in a transverse direction---the $y$ direction
in the figure---is enough to ensure that $(\partial \pi)^2$ can
remain parametrically small everywhere in the background, and thus
within the effective theory.
Note that these pathologies only occur in backgrounds where $(\partial_{\mu}\pi)^{2}$ passes through zero and goes negative---as long as $(\partial_{\mu}\pi)^{2}>0$, we can always use $\pi$ to define a single-valued time-like coordinate.

\begin{figure}[t!]
\begin{center}
\includegraphics[width=10cm]{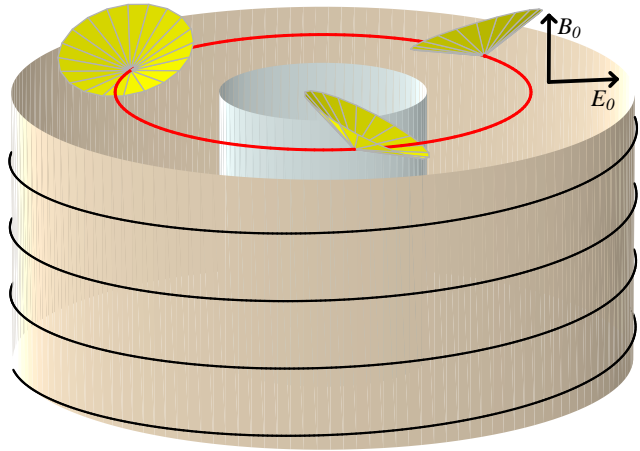}
\end{center}
\caption{\label{capacitor}
\footnotesize The field between the plates of a charged capacitor coaxial with a current-carrying solenoid is of the form $\vec{E}=\frac{A}{r}\hat{r}$ and $\vec{B}=B\hat{z}$. When $c_{1}<0$, small fluctuations at fixed $r$ propagate superluminally.  For sufficiently large fieldstrengths, but still within the regime of validity of the effective field theory, the ``causal cone'' of small fluctuations dips below horizontal, allowing for purely spacelike evolution all the way around the capacitor at fixed $t$, a dramatic violation of locality and causality.}
\end{figure}

Another particularly nice example of such closed timelike trajectories involves
the propagation of light in a non-trivial background of our ``wrong''-signed Euler-Heisenberg system in eq.~(\ref{EH}).
Consider a homogenous, static electromagnetic field with $|\vec{E}|=|\vec{B}|$ and $\vec{E}
\cdot \vec{B}=0$, such as might be found deep inside a cylindrical
capacitor coaxial with a current-carrying solenoid, as depicted in fig.
\ref{capacitor}.  Photons in this background moving orthogonal to
the field, $\vec{k}\propto \vec{E}\times\vec{B}$, and polarized
along $\vec{E}$, move with velocity
$$
v=\frac{1- c_{1} \frac{32}{\Lambda^{4}} |\vec{E}|^2}{1+ c_{1}\frac{32}{\Lambda^{4}} |\vec{E}|^2}
$$
in the direction parallel to the current and $v=1$
in the other.  If $c_{1}<0$, photons in this system propagate superluminally.
Moreover, as $c_1 \frac{32}{\Lambda^{4}} |\vec{E}|^2\to -1$, the velocity of small fluctuations diverges as their kinetic term vanishes: this is the critical value of $E$ for which the light cone of the effective metric at each point
becomes tangent to the constant time slices of an observer at rest with respect to the solenoid.
Finally, for $c_1 \frac{32}{\Lambda^{4}} |\vec{E}|^2 < -1$ the forward light cone for the effective metric
overlaps with the past of the static observer. In particular the cylinder's angular direction is at each point within
the forward effective light cone, so that a circle between the cylindrical plates at fixed Lorentz time represents a CTC for the effective metric!   Note that this configuration remains entirely within the effective theory, for while $\vec{E},\vec{B}\sim\Lambda^{2}$, all local Lorentz invariants are small---indeed they are fine-tuned to vanish.  Furthermore, the small fluctuations needed to probe these CTCs remain within the effective regime as long as their wavelengths remain large compared to $\Lambda^{-1}$.  As in the Goldstone example, violations of positivity lead to superluminality and macroscopic violations of causality.

Note that we have been tacitly working with a single positivity-violating field.  The situation is just as bad, and in some sense rather worse, if we include additional fields.  In particular, we have relied heavily on the existence, for every configuration within the regime of validity of the effective theory, of a locally comoving frame in which the condensate is at rest, i.e.~a frame in which all superluminal fluctuations propagate strictly forward in the local time-like coordinate.  If we have two superluminal fields, this is generically impossible.

Notice that attempting to define a global notion of causality, and a corresponding local Hamiltonian flow, by working in a non-inertial frame---i.e.~by working with the metric $G$ in our intrisically flat spacetime---runs into problems when the non-inertial metric admits CTCs, since the affine parameter cannot be globally defined, so evolution is a globally constrained problem.  Now, in GR, with asymptotically flat space, CTC's do not arise as long as the energy momentum tensor satisfies the null energy condition, i.e.~if the matter action satisfies certain restrictions.
It is a remarkable fact that if the matter dynamics do not feature either instabilities or superluminal modes
then the energy momentum tensor satisfies the null energy condition~\cite{fluids}. Conversely, as soon as superluminal modes are allowed, the null energy condition is lost, even in the absence of instabilities within the matter dynamics \cite{fluids}, and CTC's can in principle appear with respect to the gravitational metric $g_{\mu\nu}$ as well. Therefore, whether gravity is dynamical or not, superluminal propagation generally leads to a global breakdown of causality.

Another well-known energy condition closely related to
superluminality is the dominant energy condition.
It states that $T_{\mu\nu}t^\nu$ should a be future directed time-like
vector for any future directed time-like
$t^\mu$, i.e., there should be no energy-momentum flow outside the light-cone for any observer. 
This condition is trivially violated by a negative
cosmological constant, as well as negative tension objects such as orientifold planes in string theory. To make it meaningful one must  assume that the vacuum contribution is subtracted from $T_{\mu\nu}$.
In this form the dominant energy condition follows from the absence
of superluminality
for a large class of systems. For instance, the sound velocity in a
fluid is given by
$d p/d\rho$, and the dominant energy condition $p<\rho$ follows from
the absence of superluminality
$d p/d\rho<1$.  For a single derivatively coupled scalar field the absence of
superluminality for a general background requires the lagrangian to
be a convex function of $X \equiv (\partial_\mu\pi)^2$, ${\cal L}''(X)\geq 0$.
This is not the same as the dominant energy condition, which requires  ${\cal L}'(X)\cdot X -{\cal L}(X)>0$. For small fluctuations around the trivial background with $X = 0$, these conditions agree, but for a general background, the absence  of superluminality is a stronger condition. Thus the  absence of superluminality is a more direct and fundamental
requirement than the dominant energy condition.


\subsection{The Fate of Fate}

What have we learned about physics in a Lorentz invariant theory which allows superluminal propagation only around non-trivial backgrounds?  First, there is no Lorentz-invariant notion of causality.  Second, for observers in relative motion, disagreements about time ordering can be traced to sharp violations of locality; in sufficiently simple backgrounds, both of these complications can be avoided by a judicious choice of frame in which evolution is everywhere local and causal.  Third, in more general backgrounds, attempting to foliate spacetime into (perhaps non-inertial) constant-time slices is obstructed by the existence of closed time-like trajectories, so that time-evolution can never be locally defined but is always globally constrained.

Does this mean that effective theories which violate positivity are impossible to realize in nature?
Not necessarily.
Rather, since positivity-violating effective Lagrangians can in principle be reconstructed from experiments in completely sensible backgrounds, e.g.~by measuring low-energy scattering amplitudes in well-behaved backgrounds,
these phenomena can be interpreted as signaling the breakdown of the effective theory in pathological backgrounds.
This is a novel constraint on effective field theories, which are normally thought to be
self-consistent as long as all local Lorentz-invariants remain below a {\it UV} cutoff, so that UV-sensitive higher-dimension operators in the Lagrangian remain negligible---instead, these effective theories break down in the {\it IR} when local Lorentz-invariants get sufficiently {\it small}.  An underlying theory could complete the IR physics in two distinct ways. One possibility is that the theory simply does not admit backgrounds where local Lorentz invariants can get arbitrarlily small---for instance, if the action contains terms with inverse powers of $(\partial \pi)^{2}$. This means that even the vacuum must spontaneously break Lorentz invariance, though local physics need not be violated. Another possibility is that the underlying theory
is fundamentally non-local and capable of manifesting
this non-locality at arbitrarily large scales, while remaining Lorentz invariant.
In both cases,  positivity provides an IR obstruction to a purely UV completion of such effective theories.
Of course, no known well-defined theories, e.g.~local quantum field theories or perturbative string theories, realize such macroscopic non-locality, so positivity provides an obstruction to embedding these effective field theories into quantum field or string theory.
Any experimental observation of a violation of positivity would thus provide spectacular evidence that one of our most fundamental assumptions about Nature---macroscopic locality---is simply wrong.



\section{Analyticity and positivity constraints}

\begin{figure}[t!]
\begin{center}
\includegraphics[width=11cm]{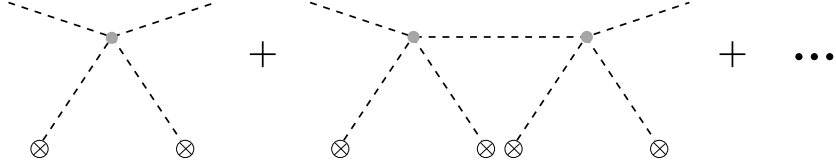}
\caption{\label{background}
\footnotesize Propagation of a small fluctuation around a background represented as a sequence of scattering events.}
\end{center}
\end{figure}

Interestingly, the UV origins of the IR pathologies we have found are visible already at the
level of 2 $\to $ 2 scattering amplitudes: with the wrong signs, these
amplitudes fail to satisfy the standard analyticity axioms of $S$-matrix theory.
To see why the UV properties of $2 \to 2$ scattering are relevant to
superluminal propagation, it is illuminating to interpret the
propagation of a fluctuation on top of a background as a scattering
process. The effect we have described corresponds to the re-summation
of all tree-level graphs depicted in fig.~\ref{background}.
That the leading vertex is a derivative interaction implies
a theoretical uncertainty of order $1/\Lambda$ on the position of the
interaction, or equivalently on the position at which our
fluctuation emerges after having interacted with the background.
This is because the derivative involves knowing the field at two arbitrarily
close points, but the closest we can take two points in the effective
theory is a distance of order $1/\Lambda$---the exact position is
fixed by the microscopic UV theory. In a typical collision,
any advance or retardation due to physics on scales smaller than the
cutoff is thus unmeasurable in the low-energy effective
theory.  However, during propagation in a translationally-invariant
background, many scattering events take place, each contributing the
same super- or sub- luminal shift. Over large distances and after
many scatterings, these small shifts add up to give a macroscopic
time advance or delay that {\it can} be measured in the effective
theory. This consideration makes it clear that the presence/absence
of superluminal excitations is a UV question: it depends on the signs
of non-renormalizable operators precisely because these interactions
cannot be extrapolated down to arbitrarily short scales.

In a local quantum field theory, the subluminality of the speed of small
fluctuations around translationally invariant backgrounds follows
straightforwardly from the fact that local operators commute
outside the lightcone. Recall that, in a free field theory, while
$\langle {\rm vac} | T(\phi(x) \phi(y)) |{\rm vac}\rangle$ is the Feynman
propagator, $\langle {\rm vac} |[\phi(x) , \phi(y)]|{\rm vac} \rangle$
determines the retarded and advanced Green's functions as
\begin{equation}
\langle {\rm vac} \left| \left[\phi(x),\phi(y)\right] \right| {\rm vac}
\rangle = D_{\rm ret}(x - y) - D_{\rm adv}(x - y) \; .
\end{equation}
Therefore, the vanishing of the commutator as an operator statement,
\begin{equation}
\left[\phi(x),\phi(y)\right] = 0 ~~~~ i\!f ~~~~ (x-y)^{2} < 0\; ,
\end{equation}
implies that $D_{\rm ret}(x - y)$ vanishes outside the lightcone.

Exactly the same logic holds in the interacting theory. The scalar
particles are interpolated by some operator $O(x)$ in the full
theory. The Fourier transform of $\langle {\rm vac} | O(x) O(y) | {\rm vac}
\rangle$ has a delta function singularity on the mass shell in
momentum space, and the pole structure is such that $\langle {\rm vac}
|\left[O(x),O(y)\right] | {\rm vac} \rangle$ is interpreted as
$D_{\rm ret}(x- y) - D_{\rm adv}(x - y)$, so that $D_{\rm ret}(x - y)$
vanishes outside the lightcone since the operator
$\left[O(x),O(y)\right]$ does. But exactly the same conclusion
follows for {\it any} translationally invariant background $|B \rangle$ of the
theory. Indeed,
\begin{equation}
\langle B \left| \left[O(x),O(y)\right] \right| B \rangle \equiv
D_{\rm ret}^B(x - y) - D_{\rm adv}^B(x - y) \; ,
\end{equation}
where $D^B(x - y)$ represents the propagator for small
fluctuations about the background $|B \rangle$. Thus again, $D_{\rm ret}^B(x -
y)$
vanishes outside the lightcone.

This argument may appear too quick---after all, our effective field theories
with the wrong signs for the
higher-dimension operators are local quantum field theories---what goes wrong
with the commutator argument?
The problem is precisely in the UV singularities associated with their only being
effective
theories. Due to the derivative interactions, the operator commutators aquire UV
singular terms
proportional to derivatives of delta functions localized on the light-cone.
These serve to fuzz-out the light cone on scales comparable to $1/\Lambda$.
Indeed, this is nothing but an operator translation of the argument at the end
of last section,
explaining how superluminality can arise as a result of a sequence of collisions
with the background field.
So it is crucial in the above argument that we are dealing with a UV complete
theory, with no UV divergent
terms localized on the lightcone in the commutators.

The commutator argument is convenient when we have the luxury of an
off-shell formulation as in local quantum field theories. But what
happens if the UV theory is not a local quantum field theory, for
instance if it is a perturbative string theory? The only
observable in string theory is the $S$-matrix. It is therefore
desirable to see whether the positivity constraints we are
discussing follow more generally from properties of the $S$-matrix.

Indeed, how is causality encoded in the $S$-matrix? After all, when
we only have access to the asymptotic states, it is not completely
clear how we would know whether the interactions giving rise to
scattering are causal or not. This was a vexing question to
$S$-matrix theorists, who wanted to build causality directly into
the axioms of $S$-matrix theory. In the end, there was no physically
transparent way of implementing causality; instead, all the physical
consequences of microcausality were seen to follow from the {\it
assumption} that the $S$-matrix as a function of kinematic
invariants is a real boundary value of an analytic function with
cuts (and poles associated with exactly stable particles) as
dictated by unitarity. Of course it is unsurprising that 
microlocality should be encoded in analyticity properties---the
textbook explanation for the absence of superluminal propagation in
mediums like glass relies on the analytic properties of the index of
refraction $n(\omega)$ in the complex frequency plane.

As we will show momentarily, the positivity constraints on the
interactions in the effective theories we have been discussing
follow directly from the dispersion relation and the assumed
analyticity properties of the $S$-matrix. As such, our conclusions
apply equally well to perturbative string theories, where the
$S$-matrix satisfies all the usual properties---unsurprisingly, as
the Veneziano amplitude arose in the framework of $S$-matrix theory.
It is of course elementary and long-understood that analyticity and
dispersion relations often imply positivity constraints (though
since such arguments are a little old-fashioned we will review them
here in detail)---what is not well appreciated is that these
positivity conditions can serve as a powerful constraint on
interesting effective field theories.

\begin{figure}[tt!]
\begin{center}
\includegraphics[height=5.4cm]{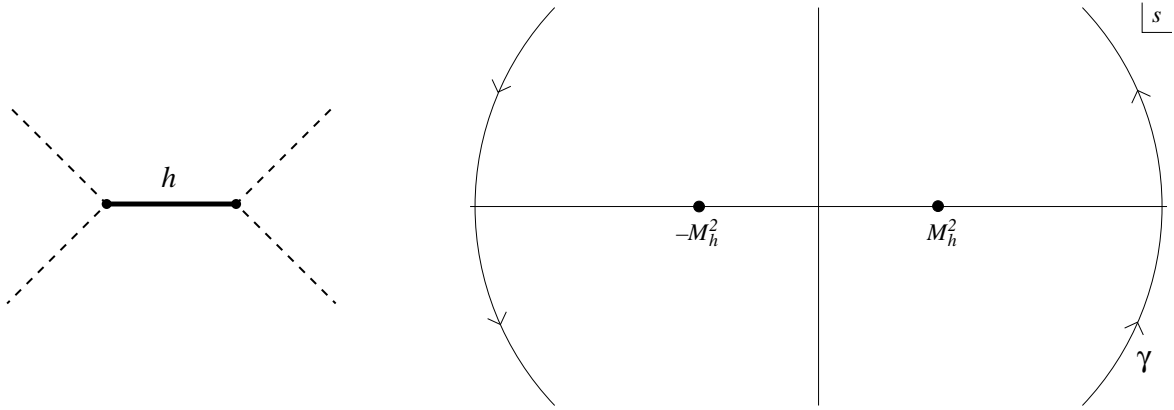}
\caption{\label{example_higgs} \footnotesize Analytic structure of the
forward $2\to2$ scattering amplitude at tree level, in the theory of
a Goldstone boson UV completed into a linear sigma model with Higgs
mass $M_h^2$. The poles arise from tree-level Higgs exchange}
\end{center}
\end{figure}

\begin{figure}[tt!]
\begin{center}
\includegraphics[height=5.4cm]{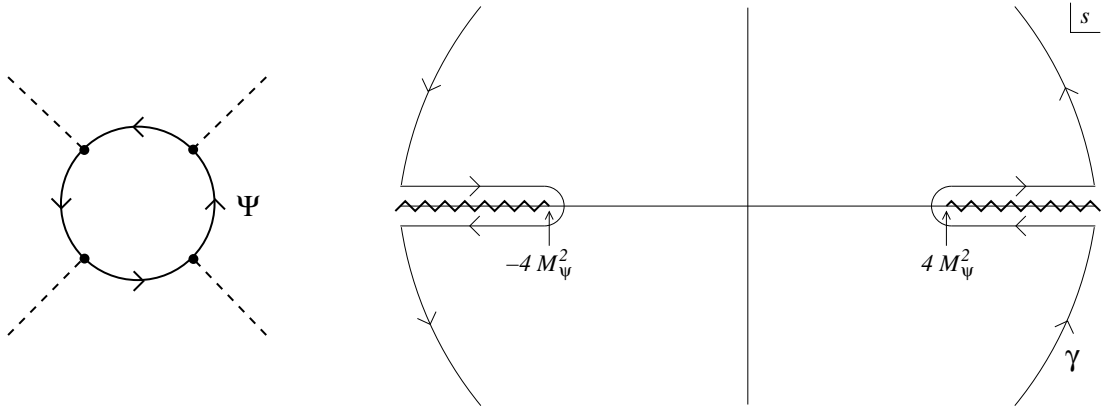}
\caption{\label{example_psi} \footnotesize Analytic structure of the
forward $2\to2$ scattering amplitude at tree level, with the
Goldstone couplings arising from integrating out a Fermion at
1-loop. The cuts starting at $s = \pm 4 M_\Psi^2$ correspond to
$\Psi$ pair production.}
\end{center}
\end{figure}

As a warm-up, let us understand why the coefficient of $(\partial
\pi)^4$ came out positive in two of our explicit examples---
integrating out the Higgs at tree-level or fermions at 1-loop. At
lowest order in the couplings the relevant diagrams are those
depicted in fig.~\ref{example_higgs} and \ref{example_psi}.
Let's consider the amplitude for $2 \to 2$ scattering, ${\cal M}(s,t)$. At
leading order and at low-energies, ${\cal M}$ is
\begin{equation}
{\cal M}(s,t) = \frac{c_3}{\Lambda^4} (s^2 + t^2 + u^2) + \dots \; ,
\end{equation}
where $u = -s - t$.
Of course this amplitude violates unitarity at energies far above
$\Lambda$, and the theory needs a UV completion.

Consider first the case where the theory is UV completed into a
linear sigma model; the full amplitude at tree level is instead
\begin{equation} \label{full_M}
{\cal M}(s,t)= \frac{\lambda}{M_h^2}\left[\frac{-s^2}{s - M_h^2} +
\frac{-t^2}{t - M_h^2} + \frac{-u^2}{u - M_h^2} \right] \; ,
\end{equation}
and of course as $s,t \to \infty$, ${\cal M}(s,t) \to$ const. Let's
further look at the amplitude in the forward direction, as $t \to
0$, and define ${\cal A}(s) = {\cal M}(s,t \to 0)$; note by crossing
symmetry ${\cal A}(s) = {\cal A}(-s)$. The analytic structure of
this amplitude in the complex $s$ plane is shown in fig. 6. Now
consider the contour integral around the contour shown in the figure
\begin{equation}
I = \oint_\gamma \, \frac{d s}{2 \pi i} \frac {{\cal A}(s)}{s^3} \,
.
\end{equation}

In the full theory, this amplitude has poles at $s = \pm M_h^2$ from
the $s$ and $u$ channel Higgs exchange. ${\cal A}(s)$ is bounded by
a constant at infinity---more generally, as long as ${\cal A}(s)$ is
bounded by $|{\cal A}(s)| < |s|^2$ at infinity, $I = 0$. On the
other hand, $I$ is equal to the sum of the residues of ${\cal A}(s)/s^3$ at its poles. Since ${\cal A}(s)/s^3 = (c_3/\Lambda^4) \, s^{-1}$ near the origin, there will be a contribution from a pole at
the origin, as well as from the poles at $s = \pm M_h^2$. Thus,
\begin{equation}
0 = I = \frac{c_3}{\Lambda^4} + 2 \, \frac{{\rm res} {\cal A}(s =
M_h^2)}{(M_h^2)^3}
\end{equation}
where the factor of $2$ accounts for the pole at $s = -M_h^2$ since
${\cal A}(s)$ is even in $s$. 
In the simple example at hand the residue of ${\cal A}$ at $s=M^2_h$ is manifestly negative from eq.~(\ref{full_M}), and so $c_3$ must be positive. However for the purpose of the future discussion
it is useful to trace how positivity of $c_3$ arises more generally from unitarity.
Indeed, as $s \to M_h^2$,
\be
{\cal A}(s)
\to \frac{{\rm res}[{\cal A}(s = M_h^2)]}{s - M_h^2 + i \epsilon} \quad \Rightarrow \quad {\rm Im}{\cal A}(s) = - \pi \delta(s - M_h^2) \, {\rm res}[{\cal A}(s = M_h^2)] \;.
\ee
Since by the optical theorem Im${\cal A}(s) = s \, \sigma(s)$ where
$\sigma(s)$ is the total cross section for $\pi \pi$ scattering, we
have
\begin{equation}
\frac{c_3}{\Lambda^4} = \frac 2 \pi \int ds \frac{s \sigma(s)}{s^3}
\end{equation}
which is manifestly positive since the cross section $\sigma(s)$ is
positive.

What about the case with the fermions integrated out at 1-loop? In
this case, the analytic structure is shown in fig. 7. There is a cut
beginning at $s = \pm 4 M_\Psi^2$ corresponding to $\Psi$ pair
production, and extending to $\pm \infty$. Now consider again the
contour integral $I$ around the curve shown in the figure. Again,
since ${\cal A}$ falls off sufficiently rapidly at infinity, $I$
vanishes. As before, there is a contribution to $I$ from the pole at
the origin, together with 2 $\times 1/(2 \pi i) \times$ the integral
of the discontinuity across the cut disc[${\cal A}(s)]/s^3$. By the
optical theorem this is again related to the total cross section for
$\pi \pi$ scattering, and we are led to the identical expression for
$c_3$ as above.

Of course this is not an accident. In fact the difference between
the analytic structures of these amplitudes is completely an
artefact of the lowest-order approximation. Let's consider the Higgs
theory at 1 loop. The amplitude will now have a cut going all the
way to the origin---the discontinuity across the cut reflecting the
(tree-level) low-energy $\pi \pi$ scattering cross section. The low
energy cross section grows and becomes largest in the neighborhood
of the Higgs resonance near $s = M_h^2$. At 1-loop, we see the
non-zero Higgs width $\Gamma$. As the physical region for $s$ is
reached from above (as per the $i \epsilon$ presecription), the
resonance is seen since the amplitude takes the usual Breit-Wigner
form $\propto (s - M_h^2 + i M_h \Gamma_h)^{-1}$. There is however
no pole on the first or physical sheet in the complex $s$ plane---the
expected pole at $s = M_h^2 -  i M_h \Gamma_h$ is reached by
continuing the amplitude under the cut to the second sheet. Of
course the presence of the resonance is visible on the physical
sheet---by a big bump in the discontinuity across the cut in the
vicinity of $s = M_h^2$. This analytic structure is exhibited in
fig. 8. Of course the analytic structure is the same for the full
amplitude at all orders, and the $\Psi$ theory as well. In fact,
this is the usual general structure of the forward scattering
amplitude---analytic everywhere in the complex plane, except for cuts
on the real axis (and poles associated with exactly stable
particles). Narrow resonances appear as poles on the second sheet.

\begin{figure}[tt!]
\begin{center}
\includegraphics[width=12cm]{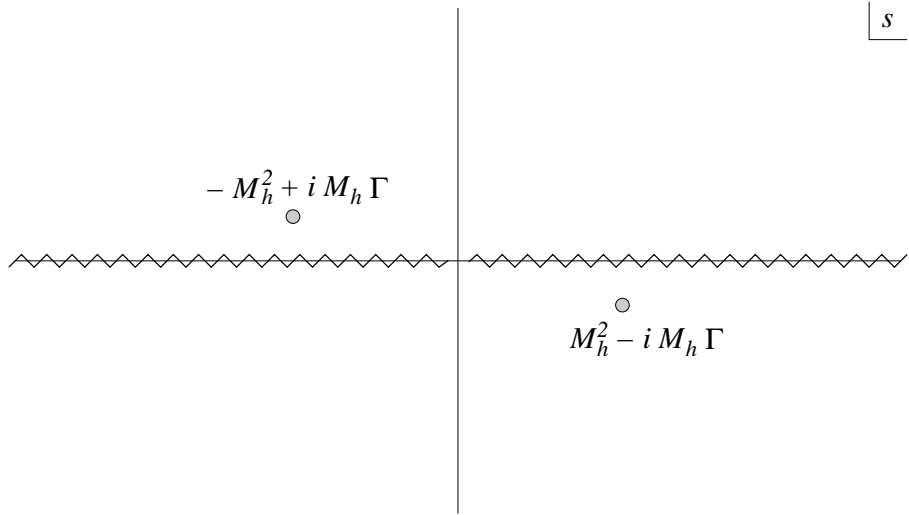}
\caption{\label{example} \footnotesize General analytic structure of
the forward $2\to2$ scattering amplitude. Poles associated with
narrow resonances are reached by going under the cut to the second
sheet.}
\end{center}
\end{figure}

Note that analyticity fixes $c_{3}$ to be {\it strictly} positive
for an interacting theory, rather than merely non-negative, as was
motivated by the IR arguments of Section 3.  Here, and in general,
the constraints coming from UV analyticity are stronger than those
observable in the effective field theory in the IR.

It is instructive to explicity see how perturbative string theory
satisfies the usual analyticity and positivity requirements. We can
also see explicitly that an analogous argument also holds for
perturbative string amplitudes. Let's consider the amplitude for
gauge boson scattering in type I string theory in 10D. At lowest
order in $g_s$, this only involves open strings, and furthermore if
we restrict the external gauge bosons to the Cartan subalgebra, the
amplitude does not have any contributions from massless gauge boson
exchnage. The scattering amplitude for gauge bosons with external
polarizations $e_{1,\cdots,4}$ in 10 dimensions has the form
\cite{joe}
\begin{equation}
{\cal M}(s, t) = g_s \, K(e_i) \left[\frac{\Gamma(-s)
\Gamma(-u)}{\Gamma(1 -s - u)} +
  \frac{\Gamma(-t) \Gamma(-u)}{\Gamma(1 -t - u)} +
  \frac{\Gamma(- s) \Gamma(-t)}{\Gamma(1 -s - t)} \right] \; ,
\end{equation}
where we are using $\alpha^\prime = 1$ units and $K$ is given by
\begin{equation}
K = -\sfrac{1}{4} \left(s \, t \, e_1 \cdot e_4 \, e_2 \cdot e_3 +
{\rm perm} \right) + \sfrac{1}{2} \left(s \, e_1 \cdot k_4 \, e_3
\cdot k_2 \, e_2 \cdot e_4 + {\rm perm} \right) \; .
\end{equation}
If we take $t \to 0$ and choose $e_{3,4}=e_{1,2}$ in order to look
at the forward amplitude relevant for the optical theorem, we find
\begin{equation}
{\cal M}(s,t \to 0) \to s \, \tan \pi s \; .
\end{equation}
This function is indeed well-behaved in the complex plane at
infinity, and is in fact bounded by $|{\cal M}(s,t \to 0)| < |s|$
away from the real axis. Thus the same arguments apply, and the
coefficient of $s^2$ in the forward amplitude is guaranteed to be
strictly positive.

Our arguments are clearly general. Other than standard analyticity
properties, all that was needed was that the forward amplitude be
bounded by $|s|^2$ at large $|s|$. In fact, under very general
assumptions, unitarity forces the high-energy amplitude in the
forward limit to be bounded by the famous Froissart bound
\cite{Froissart, Martin} ${\cal M}(s) < s \, \ln^2 s$, as follows.
As $s\to \infty$ with $t$ fixed, the total cross section is
dominated by the exchange of soft particles at large impact
parameter, so we can use the eikonal approximation to get \be {\cal
M}(s,t=-q_\perp^2)\simeq -2i~s\int d^2b ~e^{iq_\perp \cdot b}\left
(e^{2i\delta(b,s)}-1\right) . \ee Now as long as there is a mass
gap, the phase shift should fall off exponentially with impact
parameter, $\delta\sim e^{-mb} f(s) $. Locality then suggests $f(s)$
to grow no faster than a power law, $f(s)\sim s^\alpha$, with
$\alpha$ determined by the spin of the intermediate particles
(e.g.~for a single spin-$J$ particle, $\alpha= J-1$).  The forward
amplitude is thus dominated by events with $\delta=s^\alpha e^{-bm}$
of order 1, i.e.~impact parameters beneath $b<m^{-1}\ln s$, bounding
the amplitude as ${\cal M}(s) < s \, \ln^2 s$. So long as there is a
mass gap, which can often be achieved by a mild IR deformation of
the theory, a violation of the Froissart bound implies  a  dramatic
and abnormal behavior of the theory in the UV, with amplitudes that
grow faster than any  power of $s$. It thus makes sense to study the
low energy implications of a normal UV behavior which satisfies the
Froissart bound.

Let us finally give the general complete argument for positivity.
For simplicity, we restrict our attention to a general scalar field
theory with a shift symmetry $\pi \to \pi + {\rm const}$. The
leading form of the low-energy effective Lagrangian is of the form
\begin{equation}
{\cal L} = (\partial \pi)^2 + a \frac{(\partial \pi)^2
\Box \pi}{\Lambda^3} + c \frac{(\partial \pi)^4}{\Lambda^4}+
\dots
\end{equation}
Note that there is a cubic interaction term---we have not assumed a
$\pi \to - \pi$ symmetry---which might arise in a
CP violating theory for which $\pi$ is the Goldstone. As we will discuss in the next
section, the brane-bending mode of the DGP model is described
by precisely this cubic interaction.

The claim is that $c$ must be strictly positive. More precisely, we
will find a positivity constraints on the forward scattering
amplitude ${\cal A}(s) = {\cal M}(s,t \to 0)$. The argument is
virtually identical to the one used in the above examples, with two
additional technical subtleties.
First,
it is well-known that the Froissart bound can be violated by
the exchange of massless particles, such as gauge bosons and
gravitons, so we might worry that it will not hold for the
scattering of our massless $\pi$'s, which would allow amplitudes to grow too rapidly at infinity for contours to be closed.
Secondly, and relatedly, while all the non-analytic behavior of the lowest order
amplitudes of our examples was associated with UV-completion physics, the exact
amplitudes have additional cuts in the complex $s$-plane
associated to pair-production of massless particles; in the absence
of a gap, these cuts extend all the way to $s=0$.

\begin{figure}[t!]
\begin{center}
\includegraphics[width=12cm]{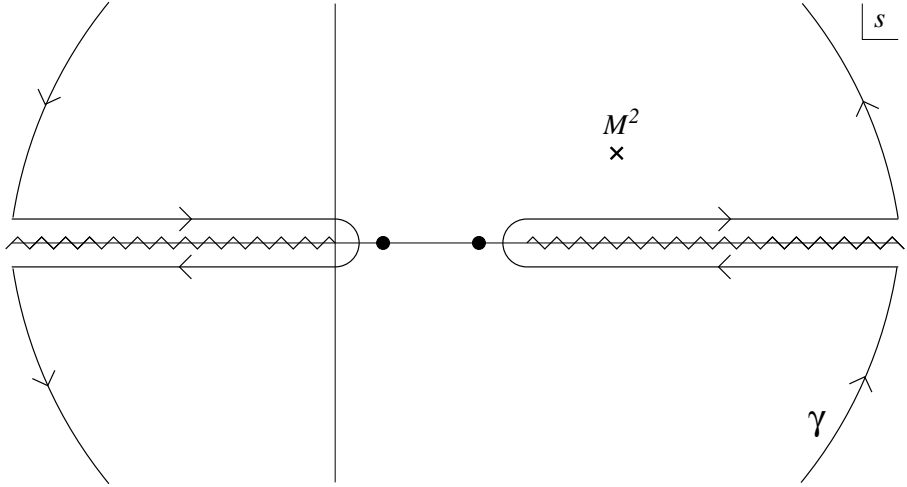}
\caption{\label{regularized} \footnotesize Analytic structure of the
forward $2\to2$ scattering amplitude in the regularized massive
theory}
\end{center}
\end{figure}

To ensure that cuts from the exchange of massless $\pi$ particles
do not modify the conclusion of positivity, we need to regulate the theory
in the IR by giving a small mass $m$ to the $\pi$ particles
(see fig.~\ref{regularized}). This also ensures that the
Froissart bound is satisfied. The $2\to2$ scattering amplitude
${\cal M}(s,t,u)$ is still symmetric in $s$, $t$, and $u$; however,
we now have $u= 4 m^2 - (s+t)$, so that the forward amplitude ${\cal A}(s) =
{\cal M}(s,0,4 m^2 - s)$ is even around the point $s= 2 m^2 $, and the
$s$-channel and $u$-channel cuts associated to $\pi$ pair production
extend on the real axis from $4 m^2$ to $+\infty$ and from $0$ to
$-\infty$, respectively (thin cuts in the figure). If the trilinear
vertex $a$ is non-zero, there is an additional contribution to the $2\to2$
scattering amplitude coming from single $\pi$ exchange, leading to
additional low-energy poles at the $\pi$ mass. However, given the
large number of derivatives involved in the leading interactions, the
residues of these IR poles scale like a positive power of $m$, and go
to zero in the massless limit. Consequently, these poles disappear in
the massless theory: in particular there is no divergence of the
amplitude in the forward ($t \to 0$) limit. This is just a
consequence of the fact that, despite the presence of massless
particles, the amplitude is dominated by short-distance
interactions.

In the forward limit, then, the $s$-channel and $u$-channel
low-energy poles are located at $s=m^2$ and at $s=3m^2$ (gray poles
in the figure), and the cuts starting from $0$ to $-\infty$ and $4
m^2$ to $+ \infty$.

Since we have modified the theory in the deep IR by adding a mass
term, we no longer want to probe the $s \to 0$ limit of ${\cal
A}(s)$, instead, we will probe the behavior of ${\cal A}(s)$ for $s$
near an intermediate scale $M^2$ with $m^2 \ll M^2 \ll \Lambda^2$.
We will do this by considering the contour integral
\begin{equation}
I = \oint_\gamma \, \frac{d s}{2 \pi i} \frac {{\cal A}(s)}{(s -
M^2)^3}
\end{equation}
Note that $M^2$ is effectively acting as an ``RG scale";  since
${\cal A}(s)$ becomes non-analytic as we approach the real axis,
this is not a convenient place to probe the amplitude, so instead we
will not put $M^2$ near the real axis but will instead consider
Re($M^2$) $\sim$ Im($M^2$).

Now, $I$ is given by the sum of the residues coming from the pole at
$s = M^2$, together with the poles near $s = m^2, 3m^2$. Since
${\cal A}(s)$ is bounded by the Froissart bound, once again
contribution to the integral from infinity can be neglected. The
contribution from the discontinuity across the cuts is determined by
the total cross section as before. We thus have
\begin{equation}
\frac 1 2 {\cal A}^{\prime \prime}(s = M^2) +\!\!  \sum_{s_* = m^2,3 m^2}
\frac{\mbox{res} {\cal A}(s = s_*)}{(s_* - M^2)^3} = \frac 1 \pi  \int _{\rm cuts} \!\! ds
\frac{s \sigma(s)}{(s - M^2)^3} 
\end{equation}
Because of the derivative interactions, the second term above is
suppressed by powers of $m^2/\Lambda^2$. Also, since at energies
beneath $\Lambda$, $\sigma(s)$ grows at least as fast as
$s^4/\Lambda^6$, for $M^2 \ll \Lambda^2$ we have that
\begin{equation}
\int _{\rm cuts} \!\! ds \frac{s \sigma(s)}{(s - M^2)^3} = 2 \int_{{\rm cut \: at \:}s>0 } \!\!\! ds \frac{s
\sigma(s)}{s^3} + {\rm corrections \, of \, order \, powers \, of\,}
\frac{M^2}{\Lambda^2}
\end{equation}
Thus we conclude that
\begin{eqnarray}
{\cal A}^{\prime \prime}(s = M^2) &=&\frac 4 \pi  \int ds \frac{s
\sigma(s)}{s^3}
+ {\cal O}\Big(\frac{M^2,m^2}{\Lambda^2}\Big) \\
&=& \rm {positive \, up \, to \, power \, suppressed \, corrections}
\end{eqnarray}
So, said precisely, the forward amplitude ${\cal A}(s)$ away from
the real axis, and for $m^2 \ll s \ll \Lambda^2$, is an analytic
function in the complex plane. Its power expansion around any point
$s_0$ in this region must begin with a term of the form $(s -
s_0)^2$ with a strictly positive coefficient.

This is all we can say in complete generality. However, in theories
where in addition to the dimensionful scale $\Lambda$ there is a
dimensionless weak coupling factor $g$ so that ${\cal M}(s,t)$ has
an expansion in $g$, we can say more. Such theories include, for
instance, weakly coupled linear sigma model completions of
non-linear sigma models, where $\Lambda$ corresponds to the Higgs
mass and $g$ is the perturbative quartic coupling in the UV theory,
or perturbative string theories, where $\Lambda$ is the string scale
$M_s$ and $g$ is the string coupling $g_s$. For $s \ll \Lambda^2$,
the tree amplitude in this theory is of the form
\begin{equation}
{\cal A}_{\rm{tree}}(s) = g \sum_{n = 1}^{\infty} c_n
\Big( \frac{s^2}{\Lambda^4} \Big)^n
\end{equation}
Note that low-energy cuts, which are absent at leading order in $g$,
appear at order $g^2$ precisely as needed for 1-loop unitarity.
Thus, by considering the contour integral
\begin{equation}
I_n = \oint_\gamma \, \frac{d s}{2 \pi i} \frac {{\cal A}(s)}{s^{2 n
+ 1}}
\end{equation}
and running through the same argument (and now ignoring the
contributions from low-energy cuts which don't exist at this order
in $g$) we conclude
\begin{equation}
c_n > 0
\end{equation}
Therefore, in a weakly coupled theory, there are an {\it infinite}
number of constraints on the effective theory: {\it the leading (in
weak coupling $g$) amplitude in the forward direction has an
expansion as a polynomial in $s^2$ with all positive coefficients}.
For example, the forward scattering amplitude in the Goldstone model is
\begin{equation}
M(s, t \to 0) = \lambda \left(\frac{s^2}{M_h^4} + \frac{s^4}{M_h^8}
+ \frac{s^6}{M_h^{12}} + \dots \right) \; ,
\end{equation}
while the amplitude for gauge boson scattering in 10D type I string
theory is
\begin{equation}
M(s, t \to 0) = g_s \left(\pi s^2 + \frac{\pi^3}{3} s^4 + \frac{2
\pi^5}{15} s^6 + \dots \right) \; ,
\end{equation}
both of which of course have all positive coefficients.


\section{The DGP Model}

The DGP model is an extremely interesting brane-world model which
modifies gravity at large distances. In addition to gravity in a
5D bulk, there is a 4D brane localized at an orbifold fixed point
with a large Einstein-Hilbert term localized on this boundary, with
an action of the form
\begin{equation}
S = 2 M_4^2 \int_{\rm brane} \! d^4 x \, \sqrt{-g} \, {\cal R}^{(4)} +
2 M_5^3 \int_{\rm bulk} \! d^4 x d y \, \sqrt{-G} \, {\cal R}^{(5)} \; ,
\end{equation}
with $M_4 \gg M_5$. The large $M_4^2$ term quasi-localizes a 4D
graviton to the brane up to distances of order $r_c \sim
M_4^2/M_5^3 \equiv 1/m$, and at larger distances gravity on the brane reverts
to being 5 dimensional.

Naively, this model makes sense as an effective field theory up to
the lower of the two Planck scales $M_5$.  However, as in the case
of massive gravity \cite{massive}, there is in fact a lower
scale
\begin{equation}
\Lambda \sim \frac{M_5^2}{M_4}
\end{equation}
at which a single 4D scalar degree of freedom
$\pi(x)$---loosely the ``brane-bending" mode---becomes strongly coupled \cite{LPR}. The classical action
for this mode can be isolated by taking a decoupling limit as
$M_4,M_5 \to \infty$, keeping $\Lambda$ fixed. In this limit
both four and five dimensional gravity are decoupled and $r_c \to
\infty$ so the physics is purely four-dimensional, leading to the effective action \cite{LPR}
\begin{equation} \label{DGP}
{\cal L} = 3 (\partial \pi)^2 - \frac{(\partial \pi)^2 \Box
  \pi}{\Lambda^3} \; .
\end{equation}
The unusual normalization of the kinetic term is for later convenience.
Note that the Lagrangian is derivatively coupled as expected for a
brane-bending mode, and that the $\pi \to - \pi$ reflection
symmetry is broken since the boundary is an orbifold fixed
point. All the interesting phenomenology of the DGP
model---including the ``self-accelerating" solution (which is
actually plagued by ghosts, as confirmed by a direct 5D
calculation in ref.~\cite{koyama2}) as well as the modification to the
lunar orbit---actually follows from this non-linear classical
Lagrangian with the scalar coupled to the trace of the energy
momentum tensor for matter fields as $(T^\mu{}_\mu/M_4) \pi$ \cite{NR}.
Indeed, the non-linear properties of this theory are what allow it to be
experimentally viable, at least classically.

Now, for realistic parameters, the scale $\Lambda$ corresponds to
$\Lambda^{-1} \sim 10^3$ km. If, at quantum level, all operators of
the form
\begin{equation}
\frac{(\partial \pi)^{2N}}{\Lambda^{4N -4}} + \dots
\end{equation}
are generated, then, despite the interesting features of the
classical theory,  the correct quantum theory would lose all
predictivity at distances beneath $10^3$ km \cite{LPR}.
It is therefore interesting to consider loop corrections in this theory,
as was initiated in \cite{LPR}, where it was shown that the tree-level
cubic term is not renormalized.  In \cite{NR}, it was shown that at loop
level only operators of the form $(\partial^2 \pi)^N$ are generated, and
with additional assumptions about the structure of the UV theory,
\cite{NR} argued that the healthy classical non-linear properties of the
theory survive quantum-mechanically.

These results all follow from the fact that
the form of the Lagrangian is preserved by a constant shift in the first derivative of $\pi$,
\be
\di_\mu \pi \to \di_\mu \pi + c_\mu \;.
\ee
Naively this suggest that any term in the Lagragian should
involve at least two derivatives on every $\pi$---however the
variation of the cubic term in eq.~(\ref{DGP}) under this
transformation is a total derivative, and therefore vanishes once
integrated. The same holds for the kinetic term, $(\di \pi)^2$.

This symmetry is nothing but 5D Galilean invariance. The position of
the brane along the fifth dimension $y_{\rm brane}(x)$ is (in some
gauge) related to the canonically normalized $\pi(x)$ by $y_{\rm
brane} = \frac{1}{m M_4}\,\pi$. The model enjoys of course full 5D
Lorentz invariance, but in the decoupling limit in which $\pi$ is
the only relevant mode, \be M_5, M_4 \to \infty \; , \qquad \Lambda
= {\rm const} \; , \ee the brane becomes flatter and flatter, the
`velocity' $\di_\mu y_{\rm brane}$ goes to zero and a 5D Lorentz
transformation acts on $y_{\rm brane}$ as a Galilean transformation.
This symmetry forces the Lagrangian to take the form \be {\cal L} =
3(\di \pi)^2 - \frac{1}{\Lambda^3} \Box \pi (\di \pi)^2 + {\cal O}
\big( \di^m (\di^2 \pi)^n \big) \; , \ee that is all further
interactions involve at least two derivatives on any $\pi$.
\footnote{Of course it is possible to make field redefinitions to
eliminate the cubic interaction term, but the theory is not free, the
tree-level $2 \to 2$ scattering amplitude is non-zero. The field
redefinition $\pi = \phi - \frac{1}{3 \Lambda^3} (\partial \phi)^2$
eliminates the DGP term but generates quartic terms
of the form $\frac{1}{\Lambda^6}(\partial \phi)^2 (\partial_\mu
\partial_\nu
\phi)^2$, as needed to reproduce the $2 \to 2$ amplitude. However,
the cubic form of the action is most convenient---first, because it
makes the Galilean symmetry simply manifest, and second, because the
coupling to matter is simple:  a linear coupling of the form $\pi
T/M_4$ to the trace of the energy momentum tensor $T$.}

Indeed, the absence of the $(\partial \pi)^{2N}$ terms is the only
thing making this effective theory special in any sense. After all, a
generic
UV theory yielding a $U(1)$ Goldstone boson $\pi$, which violates
$CP$ (and hence the $\pi \to -\pi$ symmetry), would have the same
leading cubic interaction, which is the lowest order derivative
coupling for a scalar. The only thing that can distinguish the DGP
scalar Lagrangian from a generic Goldstone theory is the presence
of the Galilean symmetry and the associated absence of $(\partial
\pi)^{2N}$ type terms in the Lagrangian. And again, it is the {\it
absence} of such $(\partial \pi)^{2N}$ terms in the effective action that
gives it a chance for non-linear health and experimental viability.

\begin{figure}[t!]
\begin{center}
\includegraphics[width=12cm]{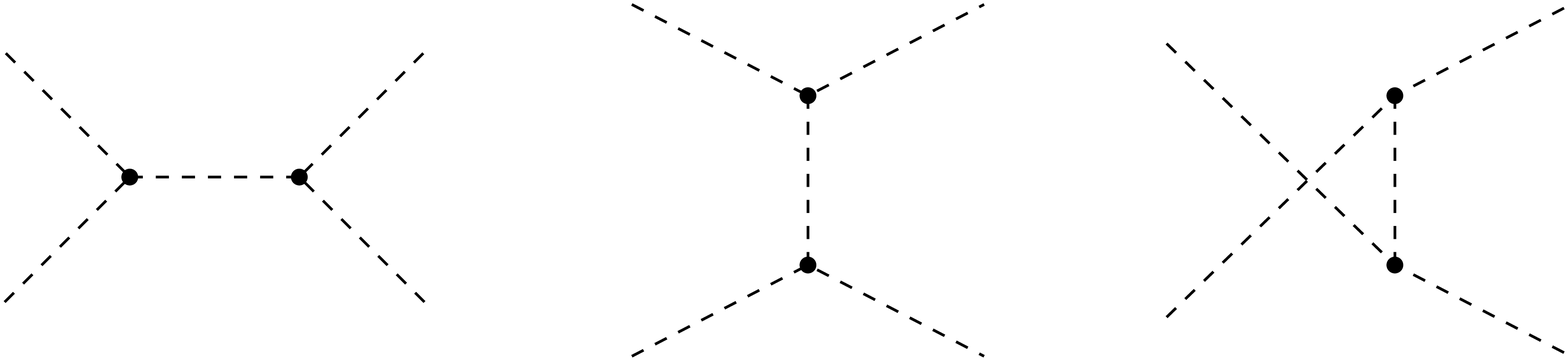}
\caption{\label{scattering}
\footnotesize Lowest order scattering amplitude in the DGP model.}
\end{center}
\end{figure}

However, {\it precisely} this property of the theory makes it
impossible to UV complete into a UV theory with usual analyticity
conditions on the $S$-matrix. As we saw in the
last section, the coefficient of the $(\partial \pi)^4$ term, which gives rise to an $s^2$ term in the forward amplitude,  must
be {\it strictly positive}. Instead, in the DGP model, this operator is forced to vanish  by the
Galilean symmetry. The amplitude for $\pi \, \pi$ scattering has a tree-level exchange contribution from the DGP term
(see fig.~\ref{scattering}) as well as contributions from the higher order term,
but they all begin at order~$s^3$
\begin{equation}
{\cal M}(s,t) = \frac{s^3 + t^3 + u^3}{\Lambda^6} + {\cal O} (s^4,t^4,u^4)
\end{equation}
In the forward limit $t \to 0$, this amplitude vanishes; and in
particular the piece proportional to $s^2$ vanishes identically.  of
course there will be {\it some} forward amplitude at even higher
orders, but these will involve even more suppression by powers of
$\Lambda$ and there will be no $s^2$ piece. We conclude that {\it it
is impossible to complete an effective theory for a scalar with a
shift symmetry of the form $\partial_\mu \pi \to
\partial_\mu \pi + c_\mu$ into a UV theory with the usual
analyticity properties for the $S$-matrix}. Again, this includes any
local quantum field theory or perturbative string theory.
Conversely, any experimental indication for the validity of the DGP
model can then be taken as the direct observation of something that
is {\it not} local QFT or string theory.

Associated with this, it is easy to see that signals about
non-trivial $\pi$ backgrounds can travel superluminally. It is trivial to
see that this is possible---the leading interaction term is cubic, and
therefore around a background, the modification of the speed of
propagation for small fluctuations is linear in the background field and
can therefore have any sign. And indeed simple physical backgrounds allow
superluminal propagation. $\pi$ is
sourced by $T$, the trace of the stress energy tensor. In the
presence of a compact spherical object of mass $M_*$, $\pi$
develops a radial background $\pi_0(r)$. The gradient of this
solution is \cite{NR}

\be \label{DGP_solution}
\pi_0' (r) = \frac {3 \Lc}{4 r} \left[ \sqrt{ r^4+\sfrac 1 {18
\pi} \, R_V^3 \, r} -  r^2\right] \; ,
\ee
where $R_V = 1/\Lambda \,(M_* / M_4)^{1/3}$ is the so-called Vainshtein radius
of the source.
In such a Schwarzschild-like solution the quadratic action for the
fluctuation $\varphi$ is \cite{NR}

\be
{\cal L}_\varphi = \Bigg[ 3 + \frac 2 \Lc \Big( \pi_0'' + \frac {2
\, \pi_0'} r \Big) \Bigg] \dot \varphi^2
- \Bigg[ 3 + \frac 4 \Lc \,\frac  {\pi_0'} r \Bigg] (\di_r
\varphi)^2 -
\Bigg[ 3 + \frac 2 \Lc \Big( \pi_0'' + \frac {\pi_0'} r \Big)
\Bigg] (\di_\Omega \varphi)^2 \; ,
\ee
where $(\di_\Omega \varphi)^2$ is the angular part of $(\vec
\nabla \varphi)^2$. The speed $c^2_{\rm rad}$ of a fluctuation
moving along the radial direction is given by the ratio between
the coefficient of $(\di_r \varphi)^2$ and that of $\dot
\varphi^2$ in the equation above; on the solution
eq.~(\ref{DGP_solution}) $c^2_{\rm rad}$ is larger than 1 for any
$r$!

\begin{figure}[t!]
\begin{center}
\includegraphics[width=9cm]{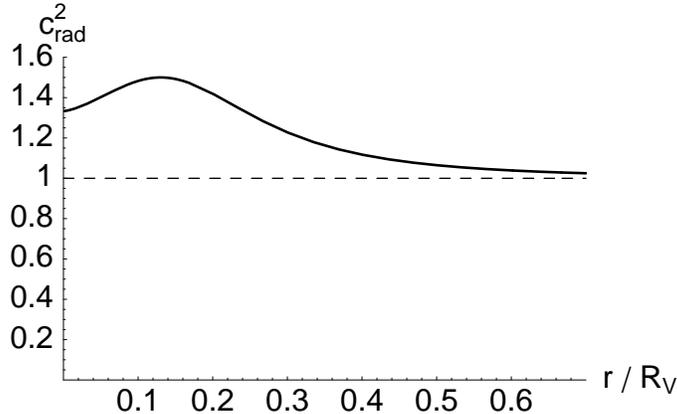}
\end{center}
\caption{\label{c2_rad}
\footnotesize The speed of radially moving fluctuations
in a Schwarzschild-like solution in DGP.}
\end{figure}

A plot of  $c^2_{\rm rad}$ versus $r$ is given in
fig.~\ref{c2_rad}: it starts from $4/3$ at $r=0$, reaches a
maximum of $3/2$ at $r \sim R_V$ and asymptotes to 1 (from above!)
for $r \to \infty$. This is an ${\cal O}(1)$ deviation from the
speed of light in an enormous region of space; for instance for
the Sun, $R_V$ is $\sim 10^{20}$ cm.
Clearly highly boosted observers can observe parametrically fast propagation,
and indeed if they boost too much they can observe the peculiar time-reversed sequence of events.
It is also easy to find spatially homogeneous and isotropic background configurations for which even observers at
rest can observe parametrically fast signal propagation.

Having found superluminal propagation, we run into the same
paradoxes as we discussed in section 2. For instance
two blobs of $\pi$ field
boosted towards each other in the $x$ direction with a small
separation in $y$ give rise to the same closed timelike curve
problems as in the two boosted blob Goldstone examples. 
However, while there we {\em assumed} the presence of suitable sources that could give rise to our paradoxical field configuration, here we expect something more. Since the simple Schwarzschild-like solution we just described features superluminal propagation, a closed timelike curve should appear in the $\pi$ field actually sourced by two masses boosted towards each other.
This is not easy to check: a quick estimate shows that in order to close the closed timelike curve the two masses must pass so close to each other that, even if their Vainshtein regions do not overlap, the presence of one mass induces sizable non-linearities close to the other, and vice versa.  In other words, the full solution is not just the linear superposition of two Schwarzschild-like solutions---new non-linear anisotropic corrections must be taken into account.
It would be interesting to further investigate such a configuration and understand
whether a closed timelike curve really arises.

It is instructive to contrast this with what happens for a generic
Goldstone theory, where the leading interaction is still the same cubic
term, but we also have the $(\partial \pi)^4$ terms. In the presence of
a generic background field $\pi_0(x)$ this interaction gives a
contribution to the quadratic Lagrangian for the fluctuations which is
linear in the background, \be \delta {\cal L} = \frac{2}{\Lambda^3}
(\di_\mu \di_\nu \pi_0 - \eta_{\mu\nu} \Box \pi_0) \, \de^\mu
\varphi \de^\nu \varphi \; . \ee

If we turn on a background with constant second derivatives, then
the field equation for the fluctuation $\varphi$ is exactly of the
form eq.~(\ref{fluct_goldstone}), with $C^\mu C^\nu$ replaced by
$\de^\mu \de^\nu \pi_0$.
Exactly as in
the DGP analysis, it appears
that superluminal signals are possible
since $\di_\mu \di_\nu
\pi_0$ has no a priori positivity property. However the $(\de \pi)^4$
term saves the day.
We can certainly set up in some
region a background with constant $\de^2 \pi_0$ and negligible
$\de \pi_0$, so that the effect of the cubic dominates over
that of $(\de \pi)^4$; but this region cannot be larger than $L
\sim \sqrt{\Lambda / \de^2 \pi_0}$, since $\de \pi_0$ grows
linearly with $x$ for constant $\de^2 \pi_0$, and after a while
the $(\de \pi)^4$ term starts dominating the kinetic Lagrangian of the fluctuations.
Once this happens, if the coefficient of $(\de
\pi)^4$ is positive there are no superluminal excitations.

The correction to the propagation speed inside the region where
the cubic dominates is $\delta c \sim \de^2 \pi_0/ \Lambda^3$,
so the maximum time advance/delay we can measure for a fluctuation
traveling all across the `superluminal region' is
\be
\delta t _{\rm max} \sim L \, \delta c \sim \frac {\de^2 \pi_0
^{1/2}}{\Lambda^{5/2}} \; .
\ee
Now, we would normally require $\de^2 \pi_0 \ll \Lambda^3$ in order for the
effective theory to make sense. In such a case we immediately get
$\delta t_{\rm max} \ll 1/\Lambda$, too small a time interval to be measured
inside the effective theory. However \cite{NR} argued that in a theory like
eq.~(\ref{DGP}) consistent assumptions about the UV physics can be made to extend the regime of validity of the effective theory
to much larger background fields and to much shorter length scales. In particular,
 in the presence of a strong background field
$\de^2 \pi_0 \gg \Lambda^3$ the effective cutoff scale is raised from $\Lambda$
to $\tilde \Lambda \sim \sqrt{\de^2 \pi_0/\Lambda}$.
In this case too the superluminal time advance is unmeasurably small: the size of the region in which
the effect of the cubic dominates over the quartic is of order of the UV effective cutoff,
$L \sim \sqrt{\Lambda / \de^2 \pi_0} \sim \tilde \Lambda^{-1}$.
In both cases the quartic saves the day.
Thus, not only does the
coefficient of the $(\partial \pi)^4$ term have to be positive, it must be
set by the same scale as the coefficient of the cubic term, a conclusion
we could have also reached from the dispersion relation arguments of the
previous section.


We have uncovered a subtle inconsistency of the DGP model. As a classical theory, it has well-defined, two-derivative, Lorentz invariant equations of motion; this  property underlies the healthy non-linear behavior of the theory and distinguishes it from more brutal modifications of gravity, such as the theory of a massive graviton.  However, just as in the simple scalar field theory examples studied in the previous sections, which also have Lorentz invariant two-derivative equations of motion, the theory suffers from a lack of a Lorentz-invariant notion of causality, which is in turn related to a violation of the usual analyticity properties of scattering amplitudes.

Of course, even in brane models respecting the usual UV locality
properties, there are DGP terms induced on the brane. What we have shown
is that we can not have a decoupling limit with $M_4/M_5 \to \infty$
holding $M_5^2/M_4$ fixed.  This suggests that there is a limit of
$M_4/M_5$ in any sensibly causal theory---it would be interesting to
investigate these questions from the geometrical perspective of the
five dimensional theory in more detail.

There is also an interesting connection between our constraint on
the DGP model and the ``weak gravity" conjecture of \cite{weakg}.
Both situations involve trying to make some interaction much weaker
than bulk gravity---in DGP it is the 4D Gravity on the brane, taking
$M_4 \gg M_5$, while in \cite{weakg}, it is the attempt to keep
$M_{Pl}$ and the cutoff of the theory fixed, but send $1/g_4^2 \to
\infty$. We have seen that a simple physical principle---requiring
subluminal signal propagation---prohibits the DGP limit. Similarly,
it appears that other general physical principles---such as the
absence of global symmetries in quantum gravity---block taking the
weak coupling limit. In both cases, there are obstacles to making
any interaction physically weaker than bulk gravity.

\section{Positivity in the Chiral Lagrangian}

There are similar positivity conditions in more familiar effective
field theories in particle physics. Consider for instance the
$SU(2)$ chiral Lagrangian, parametrized by the unitary field $U =
e^{i \pi^a \sigma^a}$,
\begin{equation}
{\cal L} = f^2 \, \mbox{tr} (\partial_\mu U^\dagger \partial^\mu U)
+ L_4 \, \big[\mbox{tr} (\partial_\mu U^\dagger \partial^\mu U)
\big]^2 + L_5 \, \big[ \mbox{tr} (\partial_\mu U^\dagger
\partial_\nu U) \big]^2 \;  + \cdots
\end{equation}

\noindent
There is a solution of the equations of motion with $\pi$ pointing
in a specifc isospin direction which we can take to be $\sigma^3$,
of the form
\begin{equation}
\pi^3(x) = c_\mu x^\mu
\end{equation}
We can look at the small fluctuations of both $\pi^3$ as well as
$\pi^{\pm}$ around this background. It is then easy to check that in
order for both $\pi^3$ and $\pi^{\pm}$ to propagate subluminally we
must have
\begin{equation}
L_{4,5} > 0
\end{equation}
In our previous Abelian examples, the 4-derivative terms were the
leading irrelevant interactions in the theory, and so did not have
any logarithmic scale dependence. On the other hand, $L_{4,5}$ {\it
are} logarithmically scale dependent; so the positivity constraint
is then actually a constraint on the running couplings at energies
parametrically smaller than $\Lambda \sim 4 \pi f$. Indeed, we can
imagine turning on a background where $\partial_\mu \pi^3$ is
approximately constant over a length scale $R$ much larger than the
cutoff scale; in order to avoid superluminality we should demand
that that running $L_{4,5}$ evaluated near this scale are positive.
Of course, the log running of $L_{4,5}$ induced off the lowest-order
2-derivative term pushes $L_{4,5}$ positive, and so in a theory
without a weak-coupling expansion, $L_{4,5}$ at low-energies are
dominated by the log running contribution and there is no
interesting constraint on the UV physics.  However in theories with
a weak coupling $g$, the matching contribution to $L_{4,5}$ at the
scale $\Lambda$ will dominate over the log-running contribution down
to energies of order $\Lambda e^{-1/g^2}$, and we can independently
identify the matching contribution to $L_{4,5}$ from the high-energy
physics from the low-energy running contribution, and hence the
positivity bound is a non-trivial constraint.

Naturally, the existence of these sorts of positivity constraints
following from dispersion theory are very well known, though not
said very explicitly; our present example was discussed (though
perhaps not widely recognized) in the literature long ago
\cite{old}.

Of course the pion chiral Lagrangian follows from QCD which is a
local quantum field theory, so these conditions must neccessarily be
satisfied. The situation is perhaps more interesting for the
electroweak chiral Lagrangian governing the dynamics of the
longitudinal components of the $W/Z$ bosons. While it is most
likely, given precision electroweak constraints, that the UV
completion involves Higgses and a linear sigma model, there may also
be more exotic possibilities, including in the extreme case a low
fundamental scale close to the electroweak scale. This physics
should manifest itself through the higher-dimension operators in the
effective Lagrangian, and assuming custodial $SU(2)$ is a good
approximate symmetry, the constraint on the electroweak chiral
Lagrangian is the same $L_{4,5} > 0$ (with the derivatives
covariantized for the $SU(2) \times U(1)$ gauge symmetry
$\partial_\mu \to D_\mu$). These operators are not associated with
the well-known constraints of precision electroweak physics---
instead, in unitary gauge $U = 1$, they represent anomalous quartic
couplings for the $W/Z$, which must be positive.

\section{Examples from String Theory}

\subsection{Little String Theory}

As we have seen, UV theories which are local or, what is the same,
satisfy the usual analyticity properties of $S$-matrix theory, give
rise to effective theories with positivity constraints on certain
leading irrelevant interactions that forbid superluminality and
macroscopic non-locality. If we experimentally measure such
interactions and find that they are zero or negative, then we have
direct evidence for a fundamentally non-local theory. But if we also
happen to know some of these operators theoretically, on other
grounds, we can use them as a locality test for the UV completion.

The prime candidate for such a test is of course $M$-theory, which
does not have a weakly coupled description, and is thought by many
to be fundamentally non-local. However, as we saw in the last
section, in gravitational theories, there is no well-defined way to
extract information about higher-dimension operators from
superluminality constraints, since the notion of the correct metric
to use for the GR lightcone can be modified by higher-dimension
operators, while just gravity already bends all signals inside the
underlying Minkowski lightcone. Associated with this, gravitational
amplitudes are dominated by long-distance graviton exchange in the
forward direction, with $t$-channel poles, and the dispersion
relation arguments can't be used.

However, we can certainly study non-gravitational UV completions of
higher-dimensional gauge theories, especially supersymmetric ones.
In five dimensions, maximally supersymmetric Yang-Mills theories are
UV completed into the {\it six} dimensional $(2,0)$ superconformal
theory, which although mysterious is still a local CFT. On the other
hand, 6D super-Yang Mills is UV completed into the 6D little string
theory, which is a non-gravitational string theory with string
tension set by the 6D Yang-Mills coupling but no small dimensionless
coupling. This is another candidate for a ``non-local" theory. This
issue can be probed if we can determine the coefficient of the $F^4$
operators in the low-energy SYM theory---if any of them have the
``wrong" sign, this would {\it prove} that the LST is dramatically
non-local.

Some of these $F^4$ terms have in fact been determined by a variety
of methods. For instance, the $U(N)$ theory has a Coulomb branch
where it is Higgsed to $U(1)^N$, and far out along the Coulomb
branch, where the $W$'s are much heavier than the little string
scale, the $F^4$ coupling between the $U(1)$'s have been computed
\cite{ofer}. They are all positive. This is perhaps not surprising,
since they are related by dualities to $F^4$ coefficients in weakly
coupled theories where the signs are determined. Thus, if LST is
really non-local, the non-locality did not put in an appearance in
the $F^4$ terms.

\subsection{Non-commutative theories}

Probably the best studied example of the Lorentz-violating system in
 string theory is a stack of parallel D-branes in flat space-time
with constant non-zero value of antisymmetric tensor field
$B_{\mu\nu}$ along the brane world-volume. It is instructive to see
how superluminality constraints work in this case. In the presence
of non-zero $B_{\mu\nu}$ open string modes localized on D-branes
propagate in the effective metric $G_{\mu\nu}$ related to the closed
string metric $\eta_{\mu\nu}$ in the following
way~\cite{Seiberg:1999vs},
\begin{equation}
\label{SWmetric}
G_{\mu\nu}=\eta_{\mu\nu}-(2\pi\alpha')^2B_{\mu\lambda}B^\lambda_\nu\;.
\end{equation}
The metric $G_{\mu\nu}$ is a direct analogue of the effective metric
discussed in sect. 3.2, so we expect its causal cone to be contained
in the light cone of the Minkowski metric. To check this let us pick
up an arbitrary light-like vector $n^\mu=(1,n^i)$ and calculate its
norm with respect to the open string metric $G_{\mu\nu}$. One finds
\[
n^\mu G_{\mu\nu}n^\nu=-(E_i-n_i E_jn^j-B_{ij}n^j)^2
\]
where $E_i=B_{0i}$. Consequently, $n^\mu$ is either space-like or
null with respect to the open string metric, implying that its
causal  cone is indeed contained in the usual light cone. Note that
this conclusion holds not only for small values of $B_{\mu\nu}$, but
in the presence of strong field as well. For the background of the
electric type one should of course require that the electric field
is smaller than the critical value when metric $G_{\mu\nu}$ changes
its singature. Physically at this point vacuum becomes unstable
towards pair production of massive charged string modes.

In the limit of large magnetic field the propagation velocity of
open modes vanishes as compared to the speed of closed modes. In
this limit closed strings decouple from open modes and dynamics of
the open sector is described by the non-commutative field theory. At
the classical level this theory exhibits approximate
Lorentz-invariance (with open string metric playing the role of the
Lorentz metric) which is broken by the higher-dimensional operators
proportional to the non-commutativity parameter. Interestingly,
these theories allow soliton solutions which can propagate faster
than the speed of ``light'' as defined by the open string metric
$G_{\mu\nu}$ \cite{Hashimoto:2000ys}. This fact however, neither
contradicts to our superluminality constraints nor leads to any
problems with causality because these solitons still propagate
inside light cone of the closed string metric which is the true
Lorentz metric of the underlying theory.

\section{Gravity}

So far, we have been discussing non-gravitational theories. In a theory
with gravity, there is a natural UV cutoff scale of order $M_{Pl}$, and it
is natural to ask whether there are constraints on $1/M_{Pl}$ suppressed
operators from our considerations.

It is easy to see that there can't be any straightforward analog of our
superluminality constraints in GR. The reason is that in a gravitational
theory, it is natural to define the light-cone by $g_{\mu \nu}$. The
effect of any higher dimension operator on the GR lightcone can then
completely be absorbed into a redefinition of the metric by $1/M_{Pl}$
suppressed operators. There is also no analog of our arguments using the
vanishing of commutators of local operators outside the light-cone. There
are no local gauge invariant operators in gravity---one of the reasons the
only observable in a quantum theory of gravity in flat space is the
$S$-matrix.

However, we might take another track.
For weak enough gravitational fields we can certainly think of
General Relativity as a Lorentz invariant theory of a spin-2 field
interacting with matter fields in Minkowski space. The underlying
Minkowski space has a metric $\eta_{\mu\nu}$ and a well defined
light-cone, which we shall refer to as the ``Minkowski
light-cone''. On the other hand a classical gravitational field
defines a new light-cone, the cone of null geodesics of the full
metric $g_{\mu\nu}$ irradiating from a point; we call this the
``gravity light-cone''. In the weak field approximation we are
considering, the fact that massless particles propagate along the
gravity light-cone takes into account the interaction of these
particles with the background gravitational field. This is very
similar to what we have been discussing so far, the propagation of
signals in a Lorentz-violating background field. Indeed we can ask
whether it is possible to turn on a gravitational field such that
the Minkowski light-cone lies inside the gravity one, so that
massless excitations, interacting with the background
gravitational field, can actually travel outside the underlying
Minkowski light-cone.

At first this seems to be trivially possible, since the effect the
gravitational field has on the dispersion relation of a massless
particle is {\em linear} in the background field itself, and as
such it has no {\em a priori} definite sign. In fact in the geometric optics limit the
dispersion relation is simply the statement that the particle
wave-vector $k_{\mu}$ is null with respect to the full
metric $g_{\mu\nu} = \eta_{\mu\nu}+h_{\mu\nu}$,
\be
(\eta^{\mu\nu}-h^{\mu\nu}) \, k_\mu k_\nu = 0 \; .
\ee If
$h_{\mu\nu}$ is not a negative-definite matrix, then there exists
a $k_\mu$ obeying the above equation that is time-like with
respect to the underlying Minkowski metric, which means that the
particle can travel outside the Minkowski light-cone. For instance
a plane gravitational wave has a non negative-definite
$h_{\mu\nu}$.

However we must wonder how we actually set up the background
gravitational field. For solving Einstein's equations we need to
fix the gauge. Since we want to preserve Lorentz invariance, we
choose De Donder gauge, $\partial^\mu (h_{\mu\nu} - \sfrac12 \, h
\, \eta_{\mu\nu}) = 0$. Einstein's equations then read
\be \Box
(h_{\mu\nu} - \sfrac12 \, h \, \eta_{\mu\nu}) = - 16 \pi G \,
T_{\mu\nu} \; ,
\ee
where $T_{\mu\nu}$ is the sources'
stress-energy tensor. Outside the sources we can further constrain
the gauge by setting $h=0$. The {\em retarded} gravitational field
ouside the sources is therefore
\be
h_{\mu\nu} (t,{\bf x})=  - 4G
\int \! d^3 r \, \frac 1 r \, T_{\mu\nu}(t-r,{\bf x} + {\bf r}) \;.
\ee
Now it is clear that $h_{\mu\nu} \, k^\mu k^\nu$ can be made
positive only if $T_{\mu\nu} \, k^\mu k^\nu$ is negative
somewhere, but this is in contradiction with the Null Energy
Condition.
Notice that a violation of the Null Energy Condition under very broad assumptions leads
either to instabilities at arbitrarily short time-scales or to superluminal propagation
in the matter sector \cite{fluids}.
The negativity of $h_{\mu\nu}$
physically means that even if gravitational waves are present, the
contribution of the static Newtonian potential to $h_{\mu\nu}$ due
to the very same sources that emitted the gravitational waves is
always larger than the oscillatory one, and negative
definite.\footnote{ At first this seems to contradict the usual
fact that far from the source the wave field decays slower than
the static one. This is indeed true at the level of
field-strenghts, but the potentials themselves both decay like
$1/r$. }

We are therefore led to the conclusion that, in the weak field
approximation we are in, it is impossible to set up a gravitational
field such that null geodesics move outside the underlying Minkowski
light-cone. Gravity `bends' all null trajectories inside the
Minkowski light-cone. This conclusion was also reached a number of
years ago in \cite{visser}.

This is a satisfying result. However, it also makes it impossible to
constrain higher-dimension operators using superluminality
arguments---since the Einstein action already pushed signals inside the
Minkowski light-cone, higher-order operators can't change this conclusion
regardless of their sign.

The difficulty of bounding higher-dimension operators suppressed by
$M_{Pl}$ is also apparent in discussing 2 $\to$ 2 scattering amplitudes.
From our previous experience, it is evident that there are positivity
constraints on $2 \to 2$ scattering in the forward direction $t \to 0$.
However, in a gravitational theory, already at tree-level, massless
1-graviton exchange dominates the $2 \to 2$ scattering amplitude as $t \to
0$,
\begin{equation}
{\cal M}(s,t) \sim G_N \frac{s^2}{t} \; ,
\end{equation}
so again, the effect of any higher order operators are swamped by the lowest order massless
1-graviton exchange which dramatically violated the Froissart bound.

In the context of a UV theory with a weak-coupling factor, however, there is some hope.
For instance, consider weakly coupled string theory, and consider the amplitude for massless closed string mode scattering.
The lowest order tree amplitude has a contribution from graviton exchange, as well as from the tower of Regge states. We can isolate the contribution from Regge states
simply by subtracting the tree-level graviton exchange diagram. The resulting subtracted amplitude then has a well-behaved limit as $t \to 0$. On the other hand, the Regge
behavior of the full amplitude as $s \to \infty$ for fixed $t$ tells us that the amplitude behaves as $s^{\alpha(t)}/t$ at large $s$, and therefore, upon subtracting the
tree-level graviton contribution that removes the $t$-channel pole, the amplitude is polynomially bounded in the complex $s$ plane. Thus, we should expect that as
$t \to 0$ the leading amplitude is of the form
\begin{equation}
{\cal M}(s,t) \to G_N \frac{s^2}{t} + \mbox{polynomial in $s^2$ with
  all positive coefficients} \; ,
\end{equation}
which again implies an infinite number of constraints on the coefficients of higher-dimension operators in the theory.

Let us see how this works explicitly for the scattering of NS-NS bosons of type II strings in 10 dimensions. The full scattering amplitude is of the form
\begin{equation}
{\cal M}(s,t,u) = - \frac{g_s^2}{4} K \frac{\Gamma(-\frac{1}{4} s)  \Gamma(-\frac{1}{4} t) \Gamma(-\frac{1}{4} u)}{\Gamma(1 + \frac{1}{4} s)
\Gamma(1 + \frac{1}{4} t) \Gamma(1 + \frac{1}{4} u)}
\end{equation}
where once again $K$ is a factor depending on external polarizations; for our purposes it is only important that as $t \to 0$, $K \sim s^4$. The amplitude can be
expanded as
\begin{equation}
{\cal M} = {\cal M}^{\rm grav} + {\cal M}^{\rm Regge} \; ,
\end{equation}
where
\begin{equation}
{\cal M}^{\rm grav} = -\frac{64 \, K }{s \, t \, u}
\end{equation}
is the tree-level graviton contribution, while $M^{\rm Regge}$ is by definition the contribution to the tree-level amplitude from the heavy Regge modes.
Of course $M^{\rm Regge}$ has a good behavior as $t \to 0$, and it is easy to check that it is indeed polynomially bounded at infinity; in fact it is bounded by $|s|^4$ at large $s$.
From our general argument, all the terms beginning with $s^4$ are guaranteed to be positive, and indeed
\begin{equation}
{\cal M}^{\rm Regge}(s, t \to 0) = - \psi_2(1) \, s^4 + \frac{- \psi_4(1)}{192} \, s^6 + \frac{- \psi_6 (1)}{92160}\,  s^8 + \dots
\end{equation}
where all the polygamma functions $\psi_n(1)$ are negative.

Thus, while we can't say anything in general about higher-order operators in gravitational theories,
there is a simple diagnostic for whether certain gravitational amplitudes do {\it not} come from a weakly coupled string theory---if the short-distance contribution
to the amplitide is a polynomial in $s^2$ with any negative coefficients, it can't come from a perturbative closed string model.


\section{Discussion}

We have shown that certain apparently consistent effective field theories described by local,
Lorentz-invariant Lagrangians are secretely non-local. The low-energy manifestation of this
lurking non-locality is the possibility of superluminal signal propagation around coherent
background fields. This creates a tension between causality and Lorentz-invariance: in such theories no Lorentz-invariant
notion of causality or locality exists.
The high-energy face of this tension is that such theories
can not be UV completed into full
theories that satisfy the usual axioms of $S$-matrix theory, specifically
the analyticity conditions that encode locality. Both local quantum field
theories and string theories satisfy these properties, so we have provided
a simple diagnostic for theories that can {\it not} be embedded into local
QFT's and string theory.

In effective theories where the particle content
or symmetries force the leading interactions to be irrelevant operators,
completely standard dispersion-relation arguments force positivity
conditions in these interactions. In terms of the scattering amplitude,
the coefficient of the leading $s^2$ term in the forward amplitude must
be strictly positive. This requirement precisely guarantees the absence of superluminal excitations around coherent backgrounds.
In weakly coupled theories, there is the stronger
constraint that the leading amplitude is a power series in $s^2$ with all
positive coefficients.

Our analysis applies to the DGP model---specifically the
four-dimensional effective theory for the ``brane-bending" mode $\pi$ that
is all that is left in a decoupling limit sending $M_4,M_5 \to
\infty$ keeping $\Lambda = M_5^2/M_4$ fixed. This scalar theory is
controlled by a nice symmetry---interpreted as Galilean invariance in the
underlying 5D theory, under which $\partial_\mu \pi \to \partial_\mu \pi
+ c_\mu$. This symmetry forces the coefficient of the leading
operator giving rise to an amplitude proportional to $s^2$ to vanish, in
contradiction to the strict positivity of this coefficient in any UV
local theory. Associated with this, signals about consistent $\pi$
backgrounds can propagate parametrically faster than light.

Similar considerations apply to the chiral Lagrangian, where the
coefficient of some of the 4-derivative terms are forced to be positive.
In the electroweak chiral Lagrangian, these turn into anomalous $W$/$Z$
quartic gauge couplings in unitary gauge. Thus, UV locality implies that
such anomalous couplings are necessarily positive.

All of our considerations in this paper have concerned
Lorentz-invariant theories with a Lorentz-invariant ground state. It
is the presence of the asymptotic Minkowski space that allows us to
get into paradoxes involving highly boosted bubbles. While our
arguments do not directly apply to theories in which the vacuum
spontaneously breaks Lorentz invariance, such as Higgs phases of
gravity \cite{ghost,Rub,Dub} or the models studied in \cite{fluids},
it would be interesting to ask whether there are any analogous
constraints to those we have discussed.

The UV face of the positivity constraints we have described are of
course implicitly well-known in dispersion theory.  In the context
of the chiral Lagrangian, precisely the positivity constraints we
describe have been discussed for instance in \cite{old}. However,
the directness with which these important signs pinpoint effective
theories that can or can not arise from local UV theories, and the
connection with superluminality and the breakdown of macroscopic
locality,  has not been appreciated. For instance, any experimental
support for the DGP model, or negative signs for anomalous quartic
gauge boson couplings, can be taken as a direct indication of the
existence of macroscopically non-local physics unlike anything ever
seen in physics, either in quantum field theory or weakly coupled
string theory. Experimental tests of positivity then provide a
powerful probe into the validity of some of our most firmly held
assumptions about fundamental physics.

Note that our results are quite likely to have
interesting connections to the physics of horizons. For example, in
theories violating positivity, a Rindler observer immersed in the
translationally invariant superluminal background can see behind the
nominal Rindler horizon to see all of Minkowski space.  Similarly, it
should be possible to see behind black hole and cosmological horizons by
tossing superluminal bubbles at them, which runs afoul of all the usual
rules about horizons and their thermodynamic properties. Another concrete
link between positivity and black-hole physics was developed in
\cite{weakg}. It would be interesting to further explore such connections.

Similarly, it would be interesting to apply positivity to various models under consideration to identify which can be embedded in local UV theories, and which are in fact fundamentally non-local.  Obvious candidates are various beyond-the-standard-models with leading derivative interactions, as well as the host of interesting varying speed-of-light models of interest in cosmology, many of which very likely run afoul of positivity.

More generally, we have studied only one very simple constraint on the low-energy effective field theory of a local, Lorentz-invariant UV-complete Quantum Field Theory deriving from the analyticity of the UV $S$-matrix, namely, positivity of forward $2\to2$ bosonic scattering amplitudes.  For example, it would be interesting to identify constraints deriving from higher scattering amplitudes, amplitudes involving fermions, gravitational amplitudes, etc.  It would also be interesting to identify other properties of UV-complete models above and beyond simple analyticity of the $S$-matrix---do they lead to interesting constraints on which effective field theories may be UV completed, and if so, what is the IR pathology of a theory which runs afoul of such a constraint?  It seems possible that there are many more constraints on just which effective theories may be consistently embedded in UV-complete theories, and, in particular, string theory.


\bigskip\noindent{\bf Acknowledgements}

\noindent We would like to thank Paolo Creminelli, Oliver DeWolfe, Jacques Distler,
Gia Dvali, Gregory Gabadadze, Leonardo Giusti, Thomas Gregoire,
Shamit Kachru, Markus Luty, Aaron Pierce,  Massimo Porrati, Valery Rubakov, Leonardo
Senatore, Misha Shifman, Raman Sundrum, Sergei Sibiryakov, Igor
Tkachev, Enrico Trincherini, and Arkady Vainshtein for helpful
conversations about these ideas. A.~A. would particularly like to
thank Aaron Pierce for invaluable assistance during the final
revisions of this note. We thank Howard Schnitzer for pointing out
elementarily incorrect things about the analytic structure of
amplitudes said in the first version of this paper, and for a
helpful conversation about it. N.~A.-H. would like to thank Lubos
Motl and especially Michal Fabinger for deconfusing him with
extremely helpful and illuminating discussions. The work of N.~A.-H.
is supported by the DOE under contract DE-FG02-91ER40654. The work
of A.~A. is supported by a Junior Fellowship from the Harvard
Society of Fellows. The work of R.~R is partly supported by the
European Commission under contract MRTN-CT-2004-503369.


\end{document}